%% file: qbt2d-honeycomb-qmc.tex
\documentclass[aps,twocolumn,prl,superscriptaddress,floats]{revtex4-2}

\usepackage{amsmath}
\usepackage{amssymb}
\usepackage{wasysym}
\usepackage{graphicx}
\usepackage{hyperref}
\usepackage{newtxtext}
\usepackage[varvw]{newtxmath}
\usepackage{mathtools}
\usepackage[shortlabels]{enumitem}
\usepackage{placeins} 
\usepackage{bbold}
\usepackage{physics}
\usepackage{multirow}

\hypersetup{
    colorlinks,
    linkcolor={blue},
    citecolor={blue},
    urlcolor={blue}
}

\graphicspath{{./}{./figs/}}

\newcommand{\rme}{\mathrm{e}}
\newcommand{\rmi}{\mathrm{i}}
\newcommand{\rmd}{\mathrm{d}}

\makeatletter
\let\MyIntOrig\int
\def\MyIntSpace{\hspace{-.25em}} 
\def\int{\MyInt}
\def\MyInt{\MyIntOrig\MyIntSkipMaybe}
\def\MyIntSkipMaybe{
  \@ifnextchar_{\MyIntSkipScript}{%
  \@ifnextchar^{\MyIntSkipScript}{%
  \@ifnextchar\limits{\MyIntSkipTok}{%
  \@ifnextchar\nolimits{\MyIntSkipTok}{%
  \MyIntSpace}}}}%
}
\def\MyIntSkipScript#1#2{#1{#2}\MyIntSkipMaybe}
\def\MyIntSkipTok#1{#1\MyIntSkipMaybe}
\makeatother

\makeatletter
\def\maketitle{
\@author@finish
\title@column\titleblock@produce
\suppressfloats[t]}
\makeatother

\usepackage{xcolor}
\usepackage[normalem]{ulem}

\begin{document}

\title{Emergent relativistic symmetry from interacting fermions on the honeycomb bilayer}

\author{Zi Hong Liu}

\affiliation{State Key Laboratory of Surface Physics and Department of Physics, Fudan University, Shanghai 200438, China}
\affiliation{Institut f\"ur Theoretische Physik and W\"urzburg-Dresden Cluster of Excellence ctd.qmat, TU Dresden, 01062 Dresden, Germany}

\author{Lukas Janssen}

\affiliation{Institut f\"ur Theoretische Physik and W\"urzburg-Dresden Cluster of Excellence ctd.qmat, TU Dresden, 01062 Dresden, Germany}

\begin{abstract}
We study the phase diagram of interacting spinless fermions on the honeycomb bilayer at charge neutrality using large-scale quantum Monte Carlo simulations.
In the noninteracting limit, the low-energy spectrum features quadratically dispersing bands that touch at the corners of the hexagonal Brillouin zone.
Weak to intermediate interactions induce a splitting of each of the quadratic band touching points into four Dirac points, located along high-symmetry directions of the reciprocal lattice.
Strong interactions lead to the formation of a layer-polarized charge density wave, which spontaneously breaks the $\mathbb Z_2$ layer inversion symmetry and opens an insulating gap in the spectrum.
We show that the semimetal-to-insulator quantum phase transition as a function of interaction is continuous and characterized by emergent relativistic symmetry. Our results for the values of the correlation-length exponent $\nu$, the order-parameter anomalous dimension $\eta_\phi$, and the fermion anomalous dimension $\eta_\psi$ agree with those of the theoretically predicted 2+1D Gross-Neveu-Ising universality class with eight two-component Dirac fermions within less than 5\%\ deviation.
We also determine the crossover scale as a function of interaction strength between the nonrelativistic semimetal state at high temperatures, characterized by dynamical critical exponent $z = 2$, and the Dirac semimetal state at intermediates temperatures, characterized by $z=1$.
Further reducing the temperature below the crossover scale at a fixed value of the interaction strength above the quantum critical point results in a classical ordering transition in the 2D Ising universality class.
\end{abstract}

\date{March 23, 2026}

\maketitle

%
In the search for novel states of matter, quantum critical points play a central role.
These are continuous phase transitions occurring at absolute zero temperature, driven by nonthermal control parameters such as pressure, doping, or twisting.
They give rise to broad quantum critical regimes at finite temperatures above the transition point, which can host nontrivial quasiparticle excitations or even admit no quasiparticle description at all~\cite{sachdevbook}.
As in classical critical phenomena, the physics in the quantum critical regime can exhibit emergent symmetries absent from the original Hamiltonian.
A well-known example studied in both classical and quantum criticality is the emergence of a U(1) symmetry~\cite{pelissetto02,lou07}. This symmetry arises, for example, in 2+1D quantum and 3D classical $q$-state clock models with $q>2$~\cite{patil21}.

In contrast to classical critical phenomena~\cite{pelissetto02}, quantum critical points can host emergent symmetries with symmetry groups containing more than one generator.
For example, a putative deconfined quantum critical point separating a N\'eel antiferromagnet and a valence bond solid is expected to host an emergent SO(5) symmetry that unifies the two order parameters~\cite{senthil23}.
Quantum critical points in Dirac materials are believed to exhibit emergent relativistic invariance~\cite{roy16,biedermann26}.
At a superconducting instability, the two-dimensional surface states of a three-dimensional topological insulator have been proposed to exhibit emergent supersymmetry, relating fermionic and bosonic quasiparticles~\cite{lee07,grover14}.
However, while the emergence of U(1) symmetry has been explicitly observed in a variety of systems~\cite{lou07,patil21,schuler23}, unambiguous evidence, beyond uncontrolled approximations, for emergent symmetries with larger symmetry groups remain scarce.
Numerical simulations of deconfined phase transitions reveal order-parameter histograms consistent with emergent SO(5) symmetry~\cite{nahum15b}; the critical exponents extracted at the putative quantum critical point, however, substantially violate rigorous bounds obtained from conformal bootstrap analyses~\cite{nakayama16, poland19}.
Simulations of the half-filled Hubbard model on the single-layer honeycomb lattice~\cite{assaad13,toldin15,otsuka16} indicate a continuous phase transition that is expected to belong to the Gross-Neveu-Heisenberg universality class with emergent relativistic symmetry~\cite{herbut06,herbut09,janssen14}. However, convergence between the critical exponents extracted from the simulations and those obtained from field-theoretical analyses has not yet been achieved, see Refs.~\cite{ladovrechis23,lang25,wang26} for recent overviews.

In this work, we investigate a model of interacting fermions on the Bernal-stacked honeycomb bilayer using large-scale quantum Monte Carlo simulations.
In the noninteracting limit, the model exhibits a nonrelativistic spectrum with quadratically dispersing bands that touch at the two inequivalent $\mathbf{K}$ points of the hexagonal Brillouin zone.
Weak to intermediate interactions split each quadratic band touching point into four Dirac points along high-symmetry directions of the reciprocal lattice.
Strong interactions drive the formation of a layer-polarized charge density wave, spontaneously breaking the $\mathbb{Z}_2$ layer-inversion symmetry and opening an insulating gap in the spectrum.
The interaction-driven semimetal-to-insulator transition is continuous, and the measured critical exponents agree with the 2+1D relativistic Gross-Neveu-Ising universality class for eight two-component Dirac fermions within 5\%, confirming emergent relativistic symmetry in the quantum critical regime.

\paragraph*{Model.}

\begin{figure}[tb!]
\includegraphics[width=0.85\linewidth]{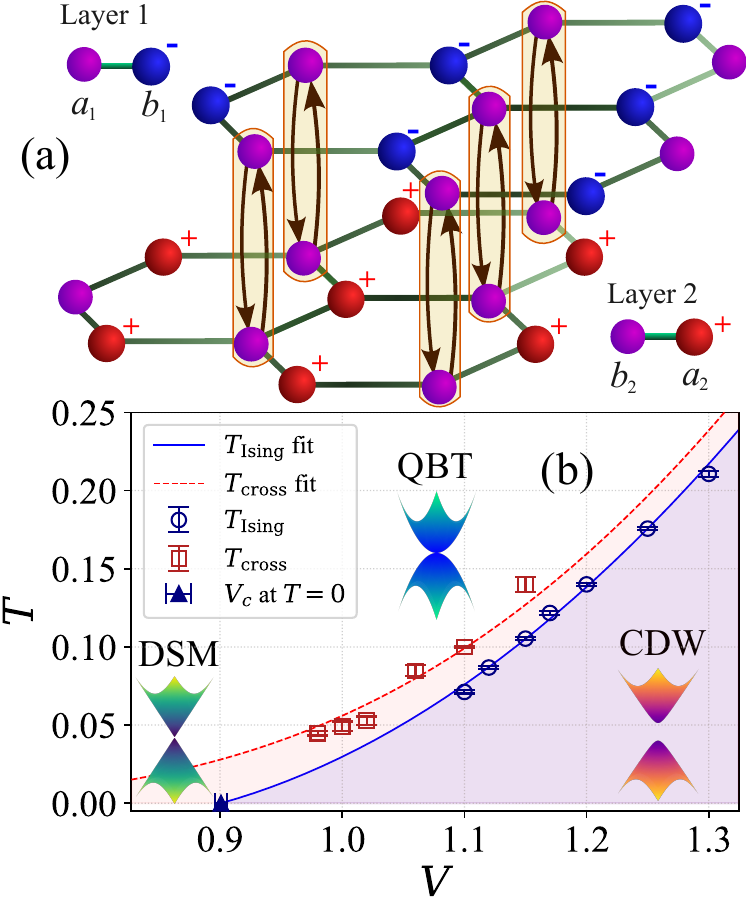}
\caption{%
(a)~Illustration of the layer-polarized charge density wave (CDW) state. Blue and red spheres denote sites with an excess and a deficit, respectively, of charge carriers relative to the average half filling. Purple-shaded regions highlight interlayer dimers hosting a single delocalized fermion. 
%
%
(b)~Phase diagram as a function of the nearest-neighbor repulsion $V$ and temperature $T$ from quantum Monte Carlo simulations.
Blue dots denote finite-temperature Ising transition points separating the CDW insulator at lower temperatures from the disordered semimetal phase at higher temperatures. Red squares mark the crossover from the Dirac semimetal (DSM) regime at intermediate temperatures to the quadratic band touching (QBT) regime at high temperatures. Insets illustrate the low-energy spectra in the three distinct regimes. Lines are guides to the eye.
%
}
\label{fig:phase-diagram}
\end{figure}

We consider a microscopic model of spinless fermions on a Bernal-stacked bilayer honeycomb lattice, described by the Hamiltonian~\cite{vafek10}
\begin{align}
H & = -t\sum_{\ell=1,2} \sum_{\langle ij\rangle} (a_{i\ell}^{\dagger}b_{j\ell}+\text{h.c.})-t_{\perp}\sum_{i}(a_{i1}^{\dagger}b_{i2}+\text{h.c.})
\nonumber \\ & \quad
+V \sum_{\ell=1,2} \sum_{\langle ij\rangle} (n_{i\ell}^{a}-\tfrac{1}{2})(n_{j\ell}^{b}-\tfrac{1}{2})
\label{eq:model}
\end{align}
where $a_{i\ell}$ and $b_{i\ell}$ are fermionic annihilation operators on the two sublattices within unit cell $i$, and $\ell = 1,2$ labeling the layer.
$t$ denotes the intralayer nearest-neighbor hopping, while $t_{\perp}$ represents the vertical interlayer hopping between sites located directly above one another.
The short-range part of the screened Coulomb repulsion acts between nearest-neighbor charge densities $n^a_{i \ell} = a_{i \ell}^\dagger a_{i \ell}$ and $n^b_{j \ell} = b^\dagger_{j \ell} b_{j \ell}$ within each layer and is controlled by the parameter $V$.
We focus on an average of two electrons per four-site unit cell, i.e., half filling, where the model exhibits particle-hole symmetry.

For $V=0$ and generic $t, t_{\perp} \neq 0$, the low-energy spectrum features two quadratically dispersing bands touching at the $\mathbf{K}$ and $\mathbf{K}'$ points of the hexagonal Brillouin zone, with the Fermi level precisely at the touching points.
The quadratic band touching (QBT) points produce a finite density of states at the Fermi level, making the system susceptible to instabilities driven by short-range interactions.
At weak to intermediate couplings, renormalization group analyses predict a splitting of each quadratic band touching point into four Dirac cones~\cite{ray18}, similar to the situation in the Hubbard model on the honeycomb bilayer~\cite{pujari16}. 
The ground state in this regime is expected to remain fully symmetric, leading to a homogeneous charge density.
By contrast, in the strong-coupling limit, the system develops a layer-polarized charge density wave (CDW) order, which breaks inversion symmetry and opens a single-particle gap~\cite{vafek10}.
The layer-polarized CDW state is illustrated in Fig.~\ref{fig:phase-diagram}(a).
In this state, pairs of sites located directly above one another form an interlayer dimer hosting a single delocalized fermion, while the remaining sites in one layer are fully occupied and those in the other layer are empty, producing a net charge imbalance between the layers.
As the interaction is reduced from the strong-coupling limit, this imbalance is expected to decrease, ultimately leading to a quantum critical point at a finite interaction strength~\cite{ray18}.

In this work, we perform large-scale determinantal quantum Monte Carlo simulations, both at zero and finite temperatures, to map out the full phase diagram of the model in Eq.~\eqref{eq:model} and to characterize the quantum critical point in an unbiased manner.
Because the hopping and interaction terms act only between different sublattices, the model is free of the sign problem thanks to Majorana reflection positivity~\cite{wei16}.
We choose units in which $t = 1$, fix $t_{\perp}=t$, and vary~$V$. 
To accelerate the simulations, we employ the submatrix update scheme~\cite{sun25}, which allows us to reach system sizes up to $4\times 27^2$ sites.

\begin{figure}[tb!]
\includegraphics[width=\linewidth]{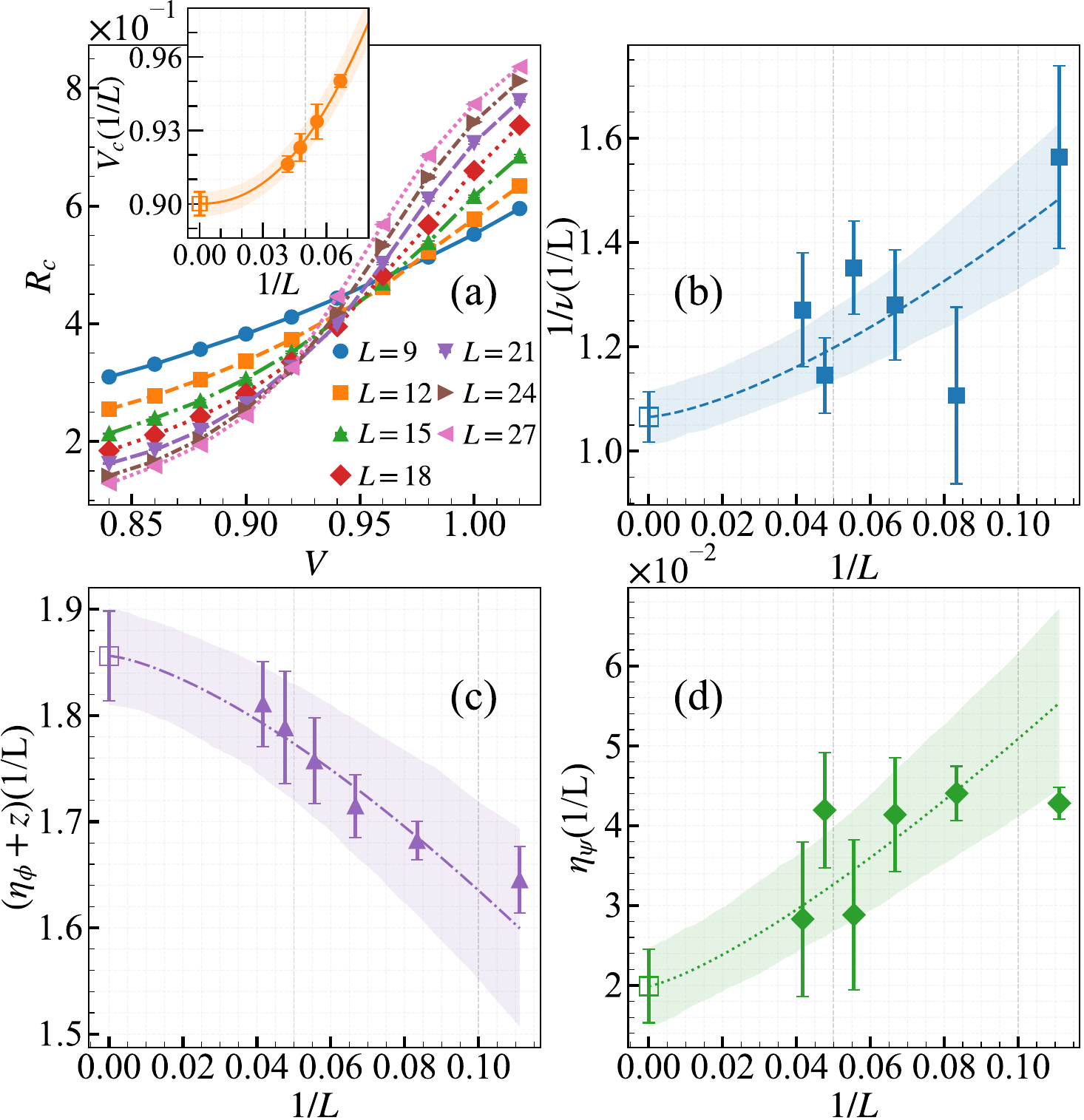}
\caption{%
(a)~Correlation ratio $R_{\mathrm c}$ as a function of the interaction $V$ at zero temperature. The inset shows the finite-size scaling behavior of the quantum critical point $V(1/L)$ obtained from the crossing of $R_\mathrm c(L)$ and $R_\mathrm c(L+3)$.
(b-c)~Crossing-point analyses of the critical exponents $1/\nu$, $\eta_{\phi}+z$, and $\eta_{\psi}$ as functions of $1/L$. The dotted curves represent the power-law scaling functions $aL^{-p}+c$.
%
}
\label{fig:exponent_t0}
\end{figure}

\paragraph*{Zero temperature.} 
%
For the ground-state phase diagram, we carry out projective quantum Monte Carlo simulations on lattices with $4\times L^2$ sites and periodic boundary conditions, with the projection length set to $\Theta=2L$ to ensure convergence.
To identify the charge-ordering phase transition, we measure the charge-correlation ratio $R_{\mathrm c}(V)=1-\Tr\chi(\mathbf k=\boldsymbol{\Gamma}+\delta \mathbf k,\tau=0)/\Tr \chi(\mathbf k=\boldsymbol{\Gamma},\tau=0)$ as a function of $V$ for different system sizes $L$.
Here, $\chi^{\alpha\beta}$ denotes the charge structure factor defined as $\chi^{\alpha\beta}(\mathbf k,\tau)=\sum_{\mathbf k} \rme^{\rmi\mathbf{k}\cdot\mathbf{r}_i} \langle n^{\alpha}(i,\tau)n^{\beta}(0,0)\rangle$ with ${\alpha,\beta}\in \{a_{i1}, b_{i1}, a_{i2}, b_{i2}\}$. The quantity $|\delta \mathbf k|\sim 2\pi/L$ denotes the smallest momentum spacing in the Brillouin zone. 
The correlation ratio is a renormalization group invariant quantity defined such that $\lim_{L\to\infty} R_{\mathrm c} = 1$ in a charge-ordered phase, while $\lim_{L\to\infty} R_{\mathrm c} = 0$ in a charge-disordered phase.
At a quantum critical point, $R_{\mathrm c}$ approaches a system-size-independent value between 0 and 1 in the thermodynamic limit.
Consequently, the crossings of the curves $R_{\mathrm c}(V)$ for different system sizes identify the critical coupling $V_{\mathrm c}$.
As shown in Fig.~\ref{fig:exponent_t0}(a), a single transition point at finite coupling, consistent with a continuous quantum phase transition between a symmetric phase for $V<V_\mathrm{c}$ and a layer-polarized CDW phase for $V>V_\mathrm{c}$, is clearly resolved.

To characterize the nature of the transition, we employ a crossing-point finite-size scaling analysis~\cite{liu22, liu24}.
Let $V_{\mathrm c}(1/L)$ denote the finite-size critical coupling defined from the crossings of $R_\mathrm{c}(V,L)$ and $R_\mathrm{c}(V,L+3)$. For increasing system size, $V_{\mathrm c}(1/L)$ approaches the true critical coupling with power-law scaling $V_{\mathrm c}(1/L)=V_{\mathrm c}+aL^{-p}$, where $a$ is a nonuniversal constant and $p$ characterizes corrections to scaling~\cite{campostrini14}. The inset of Fig.~\ref{fig:exponent_t0}(a) shows the corresponding fit, yielding $V_{\mathrm c}=0.900(5)$ in the thermodynamic limit.
%
%
Having located the transition point allows us to obtain the universal exponents that characterize the universality class of the quantum critical point.
We extract a finite-size estimate of the correlation-length exponent as $1/\nu(1/L)=\frac{1}{\ln(r)}\ln\left[ \frac{s(V_{\mathrm c}(1/L),L+3)}{s(V_{\mathrm c}(1/L),L)} \right]$, where $r=(L+3)/L$ and $s(V,L)=\partial R_{\mathrm c}(V,L)/\partial V$. The finite-size estimate approaches the true correlation-length exponent with power-law scaling $1/\nu(1/L)=1/\nu+b L^{-\omega}$, where $b$ is another nonuniversal constant and $\omega$ characterizes corrections to scaling.
Figure~\ref{fig:exponent_t0}(b) shows $1/\nu(1/L)$ as a function of $1/L$. The true correlation-length exponent is obtained from the finite-size data using Bayesian inference~\cite{harada11}, yielding $1/\nu=1.065(48)$ in the thermodynamic limit.

The anomalous dimensions $\eta_{\phi}$ and $\eta_{\psi}$ are associated with the bosonic and fermionic correlation functions, respectively. In our study, $\eta_{\phi}$ governs the scaling of the CDW order parameter, defined from the normalized density correlation at the $\boldsymbol{\Gamma}$ point, $m^2_{\text{CDW}}=\text{Tr}\chi(\boldsymbol{\Gamma},\tau=0)/L^2$. The finite-size estimate of $\eta_{\phi}$ is obtained from $\eta_{\phi}(1/L)=-z-\ln\frac{m_{\text{CDW}}^{2}(V_\mathrm{c}(1/L),L+3)}{m_{\text{CDW}}^{2}(V_\mathrm{c}(1/L),L)}/\ln r$, where $z$ corresponds to the dynamical critical exponent. We show below that the quantum critical point is characterized by $z=1$.
%
%
%
The fermionic anomalous dimension $\eta_{\psi}$ controls the scaling of the quasiparticle weight $Z_{\text{qp}}(V,L)$. Following Ref.~\cite{lang19}, we estimate $Z_{\text{qp}}(V,L)$ from the eigenvalues of the interacting Green's function near the nodal point, $G^{\alpha\beta}(\mathbf k= \mathbf K+\delta \mathbf k,\tau = 0)=\sum_{\mathbf k} \rme^{\rmi\mathbf{k}\cdot\mathbf{r}} \langle c^{\dagger}_{\alpha}(i,0)c_{\beta}(0,0)\rangle$, where $c_\alpha \in \{a,b\}$. For each $V$, we diagonalize $G^{\alpha\beta}$ and sort its eigenvalues $\{\lambda_i\}$ in descending order $\lambda_1\ge \lambda_2 \ge \lambda_3 \ge \lambda_4$. The quasiparticle weight is defined as $Z_{\text{qp}}(V,L)=\lambda_2-\lambda_3$, corresponding to the discontinuity of $n(\mathbf{k})$ at the Fermi level. The finite-size estimate of $\eta_{\psi}$ is then $\eta_{\psi}(L)=-\ln\frac{Z_{\text{qp}}(V_\mathrm{c}(1/L),L+3)}{Z_{\text{qp}}(V_\mathrm{c}(1/L),L)}/\ln r$. Figures~\ref{fig:exponent_t0}(c-d) show the Bayesian extrapolations of $\eta_{\phi}(1/L)$ and $\eta_{\psi}(1/L)$ to the thermodynamic limit $1/L=0$, giving $\eta_{\phi}=0.856(42)$ and $\eta_{\psi}=0.0199(46)$. 

The measured values of the three independent critical exponents $1/\nu$, $\eta_\phi+z$, 
and $\eta_\psi$ should be compared with theoretical expectations obtained under the assumption of relativistic invariance, which leads to Gross-Neveu-Ising criticality with eight two-component Dirac fermion flavors~\cite{ray18}.
This comparison is presented in Table~\ref{tab:exponent_compare} and reveals excellent agreement for all three observables. These results confirm that the semimetal-to-insulator transition in the present model is governed by emergent relativistic symmetry, despite the absence of a relativistic low-energy dispersion in the noninteracting spectrum. Consequently, the quantum critical point is characterized by $z=1$, which we also verify explicitly using the finite-temperature scaling analysis reported below.

\begin{table}[tb!]
\caption{%
Critical exponents from QMC simulations at the semimetal-to-insulator transition in the Bernal-stacked bilayer honeycomb model (top row, this work) in comparison with theoretical estimates for 2+1D Gross-Neveu-Ising criticality with eight two-component Dirac fermion flavors~(bottom row, reprinted from Ref.~\cite{ray18}). The excellent agreement confirms that the semimetal-to-insulator transition in the present model is governed by emergent relativistic symmetry.}
\centering 
    \begin{tabular*}{\linewidth}{@{\extracolsep{\fill} }l c c c}
    \hline\hline
     & $1/\nu$  & $\eta_{\phi}+z$ & $\eta_{\psi}$ \\[0.5ex]
    \hline
    Honeycomb bilayer (this work) & 1.065(48) & 1.856(42) & 0.0199(46) \\ 
    Gross-Neveu-Ising~\cite{ray18} & 1.018(85) & 1.868(4)\hspace{1ex}  & 0.0195(1)\hspace{1ex}  \\ 
    \hline\hline
    \end{tabular*}
    \label{tab:exponent_compare} 
\end{table}

\paragraph*{Finite temperatures.} 

\begin{figure}[tb!]
\includegraphics[width=\linewidth]{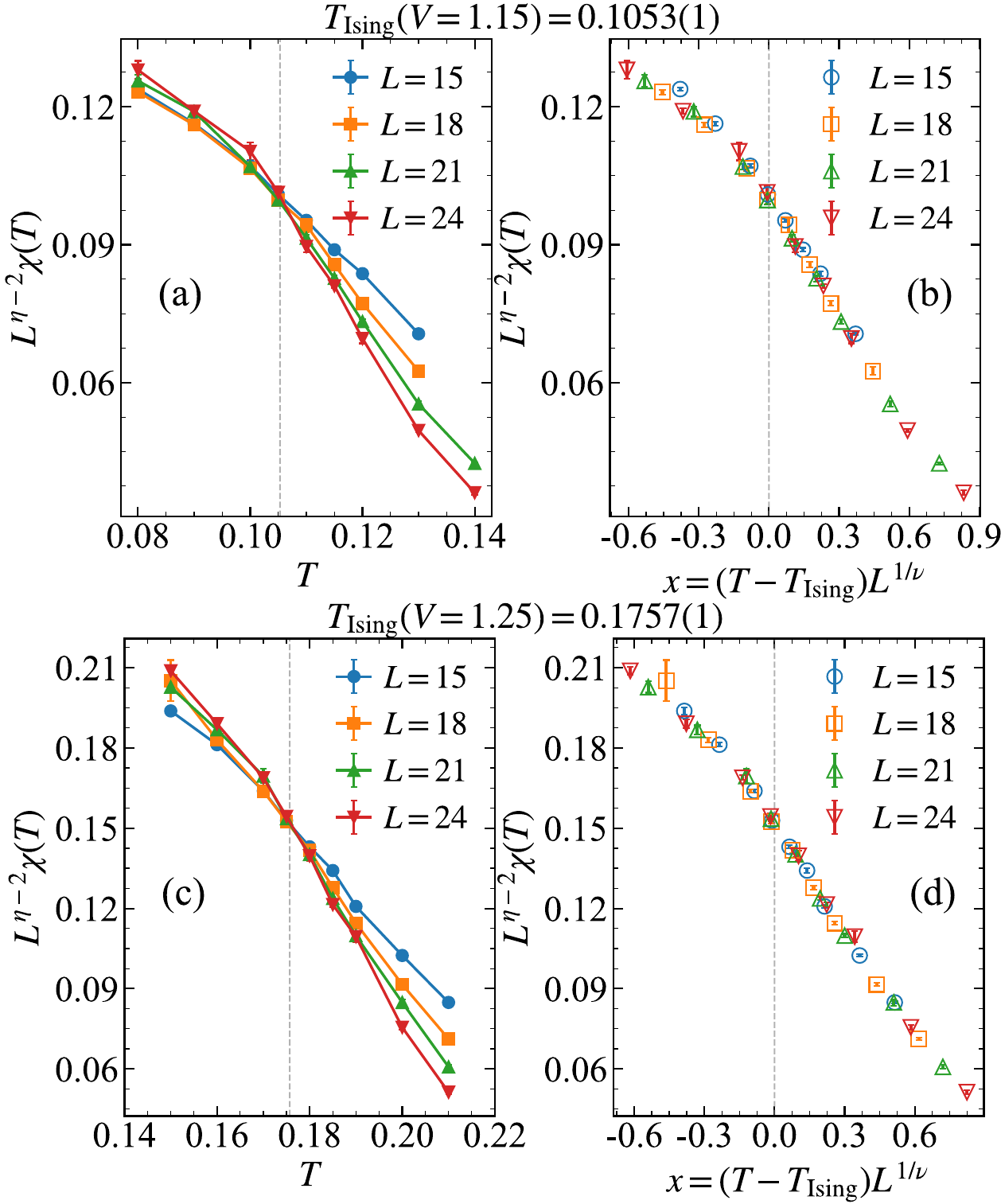}
\caption{%
(a)~Rescaled charge susceptibility $L^{\eta-2}\chi$ as a function of $T$ for fixed $V = 1.15$.
(b)~Same as (a), but plotted as a function of $(T-T_\text{Ising})L^{1/\nu_\text{Ising}}$ with $\nu_\text{Ising} = 1$.
(c,d)~Same as (a,b), but for $V=1.25$.
}
\label{fig:ft_ising_transition}
\end{figure}

As the system is heated from the low-temperature ordered regime for fixed $V>V_\mathrm c$ into the high-temperature disordered regime, two distinct characteristic temperature scales emerge. The lower scale corresponds to the melting of the CDW order, and is characterized by a two-dimensional Ising transition. The transition temperature $T_{\text{Ising}}$ is determined using the finite-size scaling form of the rescaled charge correlation function, $L^{\eta_\text{Ising}-2}\chi(T)\sim F((T-T_\text{Ising})L^{1/\nu_\text{Ising}})$ with exponents $\eta_\text{Ising}=0.25$ and $\nu_\text{Ising}=1$. As shown in Fig.~\ref{fig:ft_ising_transition}(a,c), the crossing points of $L^{\eta_\text{Ising}-2}\chi(T)$ for different system sizes identify $T_{\text{Ising}}$ at various interaction strengths. Figures~\ref{fig:ft_ising_transition}(b,d) show that the corresponding rescaled data collapse onto a universal scaling function, indicating a continuous finite-temperature transition in the 2D Ising universality class.

\begin{figure}[tb!]
\includegraphics[width=\linewidth]{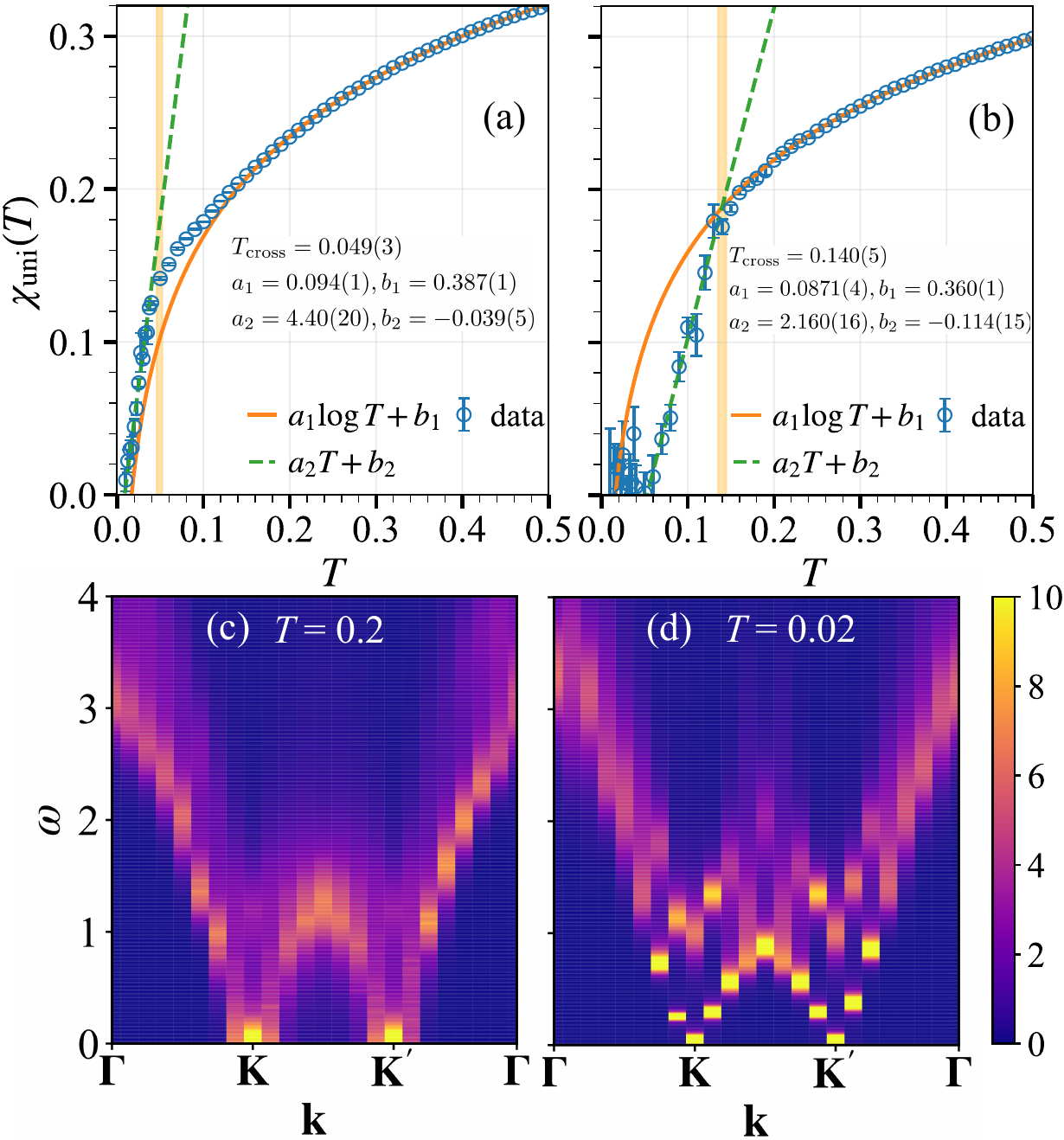}
\caption{%
(a)~Uniform charge susceptibility $\chi_{\text{uni}}$ as a function of $T$ for $V=1.00$ and $L=21$.  The shaded yellow region indicates the confidence interval of the crossover temperature $T_{\text{cross}}$, determined by the equal-residual criterion.
(b)~Same as (a), but for $V=1.15$.
(c)~Fermionic single-particle spectral function $A(\mathbf k,\omega)$ for $V=1.00$, $L=24$ and $T=0.2$.
(d)~Same as (b), but for $T= 0.02$.
%
%
}
\label{fig:ft_crossover}
\end{figure}

A second, higher characteristic temperature scale $T_{\text{cross}}$ is associated with the crossover from the relativistic regime at low and intermediate temperatures to the nonrelativistic QBT regime at higher temperatures, driven by the thermal reconstruction of the low-energy fermionic spectrum. This crossover can be detected through the uniform charge susceptibility $\chi_{\text{uni}}(T)=\sum_{\alpha\beta}\int d\tau\, \chi^{\alpha\beta}(\mathbf k=\mathbf{\Gamma},\tau)$, 
which measures U(1) charge fluctuations and is shown in Figs.~\ref{fig:ft_crossover}(a,b) for two representative values of $V$.
In the low-temperature ordered regime for $T < T_{\text{Ising}}$, $\chi_{\text{uni}}(T)$ is exponentially suppressed, reflecting the gapped single-particle spectrum in the CDW phase. For temperatures above $T_{\text{Ising}}$, we expect $\chi_{\text{uni}}(T) \propto T^{d/z-1}$~\cite{hohenadler22} for $d\neq z$, where $d=2$ is the spatial dimension and $z$ is the dynamical critical exponent, which characterizes the relative scaling of spatial and temporal correlations. For $d = z$, logarithmic corrections to the above scaling, arising from the finite density of states at the band touching point, are expected~\cite{castroneto09, supplemental}.
In the intermediate-temperature regime near and above $T_\text{Ising}$, $\chi_{\text{uni}}(T)$ exhibits linear scaling $a_2 T+b_2$, consistent with Dirac semimetallic (DSM) behavior characterized by $z=1$. At higher temperatures, $\chi_{\text{uni}}(T)$ follows a logarithmic form $a_1\log T+b_1$, characteristic of the QBT regime~\cite{castroneto09, supplemental}.
To determine the crossover scale $T_{\text{cross}}$, we employ an equal-residual criterion. The data are separately fitted in the DSM-like (linear) and QBT-like (logarithmic) temperature ranges. The residuals of $\chi_\text{uni}(T)$ with respect to each fit are then evaluated in the intermediate region, and $T_{\text{cross}}$ is identified as the temperature at which the residuals change sign, as indicated by the yellow shaded region in Figs.~\ref{fig:ft_crossover}(a,b). The resulting crossover scale $T_\text{cross}$ as a function of $V$ is shown in Fig.~\ref{fig:phase-diagram}(b) (red dots).

Additional evidence for the finite-temperature crossover between the DSM and QBT regimes is provided by the single-particle spectral function $A(\mathbf{k},\omega) =  \pi \sum_{n,\alpha}  | \langle n | {c}^{\dagger}_{\alpha}(\mathbf{k}) | 0 \rangle |^2 \delta(  E_n - E_0 - \omega)$, where $| n \rangle$ and $E_n$, $n = 0,1,2, \dots$, correspond to the energy eigenstates and eigenvalues, respectively.
Figures~\ref{fig:ft_crossover}(c,d) show $A(\boldsymbol{k},\omega)$ obtained using the stochastic analytical continuation method~\cite{alf-doc,beach04} for $V = 1.00$ at two different temperatures.
At low temperature $T=0.02 < T_\text{cross}$ [Fig.~\ref{fig:ft_crossover}(d)], the data are consistent with a linear dispersion around the $\mathbf K$ points, suggesting the formation of interaction-driven Dirac cones at low energies.
At high temperature $T=0.2 > T_\text{cross}$ [Fig.~\ref{fig:ft_crossover}(c)], the spectral weight shifts toward the Fermi level at several momenta near the $\mathbf K$ points, indicative of a quadratic band dispersion. 

\paragraph*{Discussion.}

We have studied a model of interacting fermions on the Bernal-stacked honeycomb bilayer. Despite the absence of a relativistic dispersion at the lattice scale, the system exhibits a quantum critical point characterized by emergent relativistic invariance at low energies. This behavior can be understood as arising from the interaction-induced splitting of each quadratic band touching point into four Dirac cones. While this scenario has previously been proposed in the context of simulations of the bilayer-honeycomb-lattice Hubbard model~\cite{pujari16,ray18}, the key advantage of the present spinless model is that the universal quantum critical behavior can be identified unambiguously and shown to belong to the expected Gross-Neveu-Ising universality class with eight two-component Dirac fermion flavors.

Our results call for refined transport, thermodynamic, and spectroscopic studies of Bernal-stacked bilayer graphene, to reveal the relativistic symmetry emergent at intermediate temperatures above the Mott transition. The Hall coefficient, for instance, is expected to scale with temperature as $R_{\text{H}} \propto T^{-2/z}$, with $z \simeq 2$ for $T > T_{\text{cross}}$ and $z \simeq 1$ for $T < T_{\text{cross}}$. The crossover should also manifest in Landau-level spectroscopy experiments~\cite{li09}, revealing a transition from QBT-like scaling of the Landau levels at high temperatures, $E_N \propto \pm B \sqrt{N(N-1)}$, to Dirac-like scaling at intermediate temperatures, $E_N \propto \pm \sqrt{BN}$~\cite{geim07}.

\paragraph*{Acknowledgments.}
We thank
S.~Ray
for valuable discussions and collaboration on related earlier work.
This work has been supported by the Deutsche Forschungsgemeinschaft (DFG) through SFB 1143 (A07, Project No.\ 247310070), the W\"urzburg-Dresden Cluster of Excellence \textit{ctd.qmat} (EXC 2147, Project No.\ 390858490), and the Emmy Noether program (JA2306/4-1, Project No.\ 411750675).
 The authors gratefully acknowledge the computing time made available to them on the high-performance computer at the NHR Center of TU Dresden. This center is jointly supported by the German Federal Ministry of Education and Research and the state
governments participating in the NHR~\cite{nhr-alliance}.
%




\FloatBarrier
\bibliographystyle{longapsrev4-2}
\bibliography{qbt2d-honeycomb-qmc}

\clearpage

\input{supplemental}

\end{document}

%% file: supplemental.tex

\setcounter{figure}{0}
\setcounter{equation}{0}
\setcounter{table}{0}

\makeatletter
\renewcommand{\c@secnumdepth}{0}
\makeatother

\renewcommand\theequation{S\arabic{equation}}
\renewcommand\thefigure{S\arabic{figure}}
\renewcommand\thetable{S\arabic{table}}


\title{Supplemental Material for\\ ``Emergent relativistic symmetry from interacting electrons on the honeycomb bilayer''}
\date{\today}

\begin{abstract}
The Supplemental Material contains
(i)~technical details of the quantum Monte Carlo simulations,
(ii)~additional finite-size scaling analyses,
(iii)~a discussion of the temperature dependence of the charge susceptibility in the QBT regime, and
(iv)~additional finite-temperature Monte Carlo data for the charge susceptibility.
\end{abstract}

\maketitle

\section{Quantum Monte Carlo simulations}

In this supplemental section, we provide further details of our determinantal quantum Monte Carlo simulations.
In our implementation, the expectation value of a general observable $O$ is computed differently at zero and finite temperatures.
At finite temperature $T = 1/\beta$, it is given by the thermal ensemble average
\begin{equation}
\left\langle O\right\rangle_{T>0} =\frac{1}{Z}\Tr\left\{ O\rme^{-\beta H} \right\},
\end{equation}
where $Z$ denotes the partition function.
At zero temperature, the expectation value corresponds to the ground-state average
\begin{equation}
\left\langle O\right\rangle_{T=0} =
\frac{\langle \psi_{0}|O|\psi_{0}\rangle }{\langle \psi_{0}|\psi_{0}\rangle }
=\frac{\langle \psi_\mathrm{t}| \rme^{-\Theta H} O \rme^{-\Theta H} | \psi_\mathrm{t} \rangle }{\langle \psi_\mathrm{t} | \rme^{-2\Theta H} |\psi_\mathrm{t}\rangle }\,,
\end{equation}
where $\left|\psi_{0}\right\rangle$ is the exact many-body ground state obtained by imaginary-time projection from a trial Slater determinant $\left|\psi_\mathrm{t}\right\rangle$, $\left|\psi_{0}\right\rangle=\lim_{\Theta\rightarrow \infty} \rme^{-\Theta H}\left|\psi_\mathrm{t}\right\rangle$.
In our simulations on lattices with $4 \times L^2$ sites, we set the projection length to $\Theta = 2L$, which is sufficient to ensure convergence.

To evaluate the time-evolution operator, we apply a Trotter decomposition,
\begin{equation}
\rme^{-\beta H}=\prod_{n=1}^{M} \rme^{-H_0 \Delta\tau/2 } \rme^{-H_\text{int} \Delta\tau} \rme^{-H_{0} \Delta\tau/2 }+\mathcal O(\Delta\tau^{2})\,
\end{equation}
where $\Delta\tau=\beta/M$ for finite temperatures, and similarly in the zero-temperature simulations.
Here, $H_0 = H \big |_{V \to 0}$ is the kinetic part of the Hamiltonian and $H_\text{int} = H \big|_{t,t_\perp \to 0}$ is the interaction part.
In our simulations, we employ a Trotter time step $\Delta \tau = 0.1$, for which discretization errors are negligible.

We decouple the interaction term via a continuous Hubbard-Stratonovich transformation. Using
\begin{equation}
(a_{i\ell}^{\dagger}a_{i\ell}-\tfrac{1}{2})(b_{i\ell}^{\dagger}b_{i\ell}-\tfrac{1}{2})-\tfrac{1}{4}=-\tfrac{1}{2}(a_{i \ell}^{\dagger}b_{i \ell} + \text{h.c.} )^{2}\,,
\end{equation}
we obtain 
\begin{align}
\rme^{-\Delta\tau H_\text{int}} & =
\prod_{\left\langle ij\right\rangle, \ell } \rme^{\Delta\tau V/4} \rme^{(a_{i\ell}^{\dagger}b_{i\ell} + \text{h.c.})^{2} \Delta\tau V/ 2}\nonumber \\
& =C\prod_{\left\langle ij\right\rangle, \ell }\int d\phi_{ij,\ell}e^{-\phi_{ij,\ell}^2/2} \rme^{\sqrt{\Delta\tau V}\phi_{ij,\ell}(a_{i\ell}^{\dagger}b_{i\ell}+\text{h.c.})}\,,
%
%
 %
\end{align}
where $C$ is an unimportant normalization constant. The auxiliary field $\phi_{ij,\ell}$ thus appears as a fluctuating bond variable that modulates the nearest-neighbor intralayer hopping amplitude in each Monte Carlo configuration. After Hubbard-Stratonovich decoupling, the action becomes quadratic in the fermionic degrees of freedom, which can then be integrated out exactly,
\begin{align}
Z& = \int \mathcal D[\phi] \, \rme^{-\frac12 \sum_{\left\langle ij\right\rangle, \ell}\phi_{ij,\ell}^{2}}
\nonumber \\ & \qquad
\times \det\left(1+\prod_{\ell} \rme^{-\Delta\tau T/2} \rme^{-\Delta\tau V([\phi_\ell])} \rme^{-\Delta\tau T/2}\right)\nonumber \\
 & \equiv \int \mathcal D [\phi] P\left(\phi\right)W\left(\phi\right)
\end{align}
Here $P\left(\phi\right)\coloneqq \prod_{\langle ij\rangle, \ell} \rme^{-\phi_{ij,\ell}^{2}/2}$ is the bosonic weight, and $W(\phi) \coloneqq \det(1+\prod_{\ell} \rme^{-\Delta\tau T/2} \rme^{-\Delta\tau V([\phi_\ell])} \rme^{-\Delta\tau T/2})$ is the fermion determinant, with $T$ and $V([\phi_\ell])$ being the hopping and vertex matrices.
Because the Hubbard-Stratonovich channel satisfies Majorana reflection positivity, we have $W\left(\phi\right)\ge 0$, ensuring that the model is free of the fermion sign problem.

\section{Critical exponents from data collapse}

\begin{table*}[t!]
\centering
\caption{Critical coupling $V_\mathrm c$, correlation-length exponent $1/\nu$, and anomalous dimensions $\eta_\phi$ and $\eta_\psi$ from data collapse analysis of correlation ratio $R_\mathrm c$, CDW order parameter $m_\text{CDW}^2$, and quasiparticle weight $Z_\text{qp}$ using data sets from lattices with $L \geq L_\text{min}$.
\label{tab:collapse}}
\begin{tabular*}{\linewidth}{@{\extracolsep{\fill}}c|ccc|cc|cc}
\hline \hline
$L_{\min}$ & $V_{c}$ & $1/\nu$ & fitting $\chi^{2}(R_\text c)$ &  $\eta_{\phi}+z$ & fitting $\chi^{2}(m_\text{CDW}^2)$ & $\eta_{\psi}$ & fitting $\chi^{2}(Z_\text{qp})$ \\
\hline 
12 & $0.938(1)$ & $1.31(2)$ & $16.70$ & $1.726(2)$ & $19.02$ & $0.039(1)$ & $10.03$\\
15 & $0.933(1)$ & $1.199(26)$ & $11.41$ & $1.739(3)$ & $16.44$ & $0.036(1)$ & $6.64$ \\
18 & $0.925(1)$ & $1.14(3)$ & $7.12$ & $1.761(5)$ & $15.54$ & $0.034(1)$ & $7.10$ \\
21 & $0.919(1)$ & $1.10(4)$ & $4.24$ & $1.773(7)$ & $16.70$ & $0.033(1)$ & $8.36$ \\
%
%
%
\hline\hline
\end{tabular*}
\end{table*}

\begin{figure}[tb!]
\includegraphics[width=\linewidth]{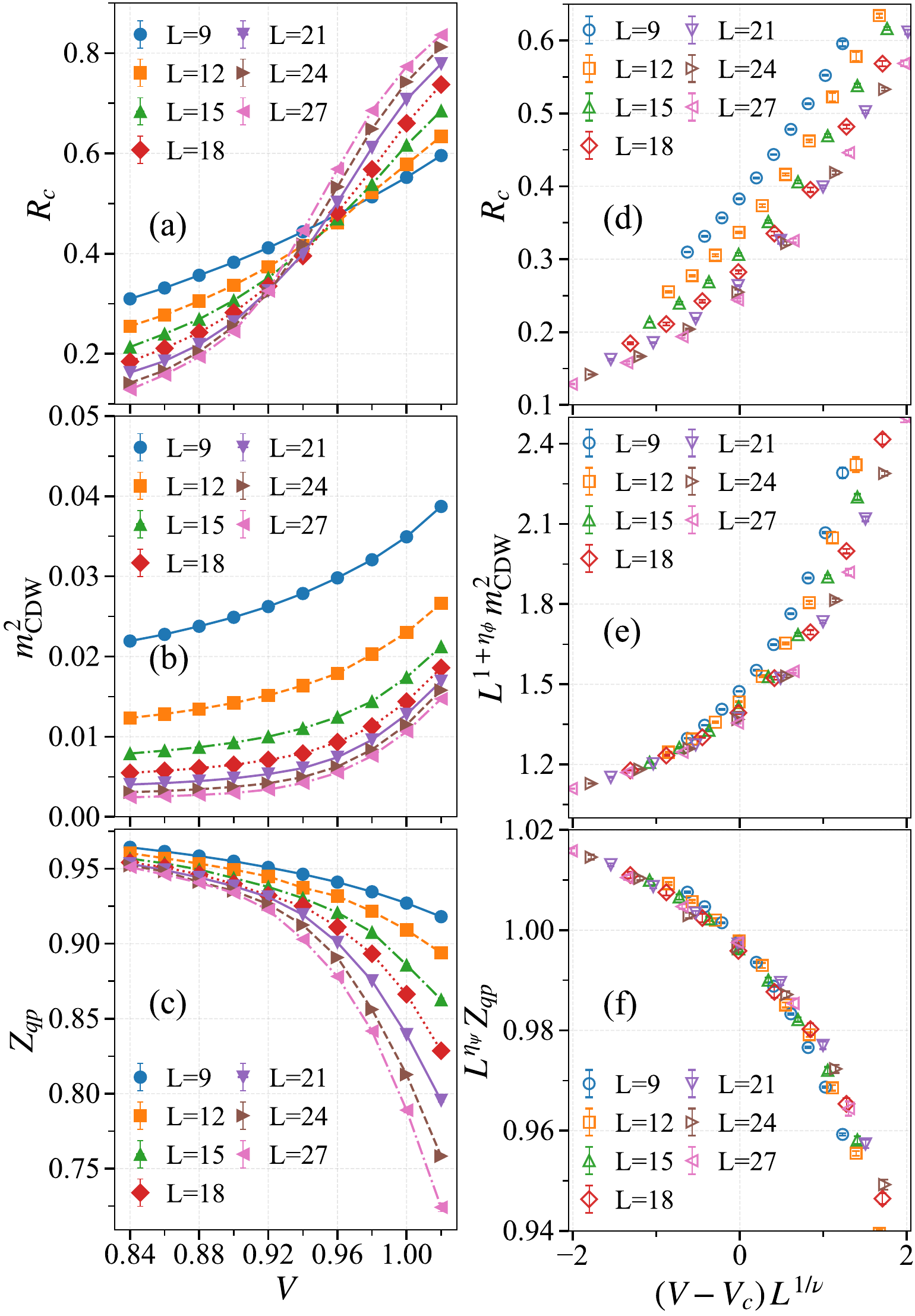}
\caption{%
(a)~Correlation ratio $R_\mathrm c$ as a function of $V$ for different system sizes. 
(b)~Same as (a), but for the order parameter $m^2_\text{CDW}$. 
(c)~Same as (a), but for the quasiparticle weight $Z_\text{qp}$.
(d)~Correlation ratio $R_\mathrm c$ as a function of the rescaled tuning parameter $(V-V_\mathrm c)L^{1/\nu}$ for different system sizes, using $V_\mathrm{c} = 0.9$ and $1/\nu = 1.06$ as obtained from the crossing-point analysis presented in the main text.
(e)~Same as (d), but for the rescaled order parameter $L^{z+\eta_\phi} m^2_\text{CDW}$, using $z+\eta_\phi = 1.85$.
(f)~Same as (d),  but for the quasiparticle weight $L^{\eta_{\psi}}Z_\text{qp}$, using $\eta_\psi = 0.02$.
%
%
\label{fig:raw_and_rescale_data_t0}}
\end{figure}

\begin{figure}[tb!]
\includegraphics[width=\linewidth]{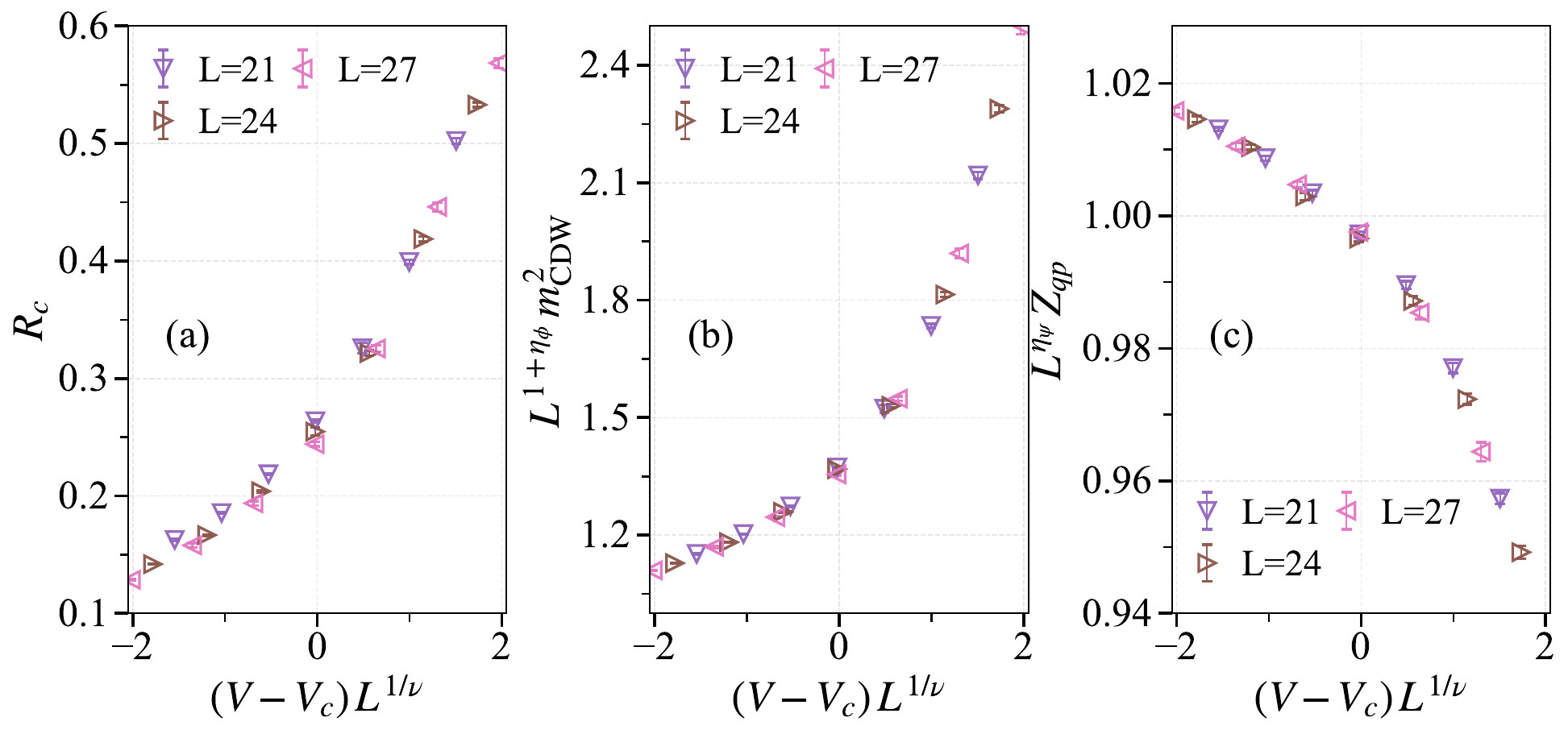}
\caption{Same as Figs.~\ref{fig:raw_and_rescale_data_t0}(d-f), but including only results for system sizes $L \geq L_\text{min}=21$, illustrating the convergence.
%
%
\label{fig:raw_and_rescale_data_t0_2}}
\end{figure}

In this supplemental section, we perform a data-collapse analysis as an independent consistency check of the critical exponents reported in the main text. Within the finite-size scaling framework~\cite{campostrini14}, the relevant observables are expected to follow universal scaling forms,
\begin{align}
R_{\mathrm{c}}(V,L) & =f_{0}^{R}(v_\mathrm{r} L^{1/\nu})+L^{-\omega}f_{1}^{R}(v_\mathrm{r} L^{1/\nu}),
\label{eq:r_scaling}
\\
m_{\text{CDW}}^{2}(V,L) & =L^{-z-\eta_{\phi}}
[f_{0}^{m}(v_\mathrm{r} L^{1/\nu})+L^{-\omega}f_{1}^{m}(v_\mathrm{r} L^{1/\nu})],
\label{eq:m_scaling}
\\
Z_{\text{qp}}(V,L) & =L^{-\eta_{\psi}}
[f_{0}^{z}(v_\mathrm{r} L^{1/\nu})+L^{-\omega}f_{1}^{z}(v_\mathrm{r} L^{1/\nu})],
\label{eq:z_scaling}
\end{align}
where $v_\mathrm{r}=V-V_c$ denotes the reduced coupling and $\omega > 0$ is a corrections-to-scaling exponent. In the thermodynamic limit, the correction terms proportional to $L^{-\omega}$ vanish and the observables collapse onto the corresponding universal scaling functions $f_{0}^{R,m,z}\left(v_\mathrm{r} L^{1/\nu}\right)$. 

For finite sizes, the correction terms lead to deviations from perfect scaling collapses. In practice, we account for these corrections by successively discarding data from small system sizes and performing the collapse using data with $L\ge L_{\text{min}}$. To perform the collapse without the knowledge of the explicit form of the universal scaling functions $f_{0}^{R,m,z}\left(v_\mathrm{r} L^{1/\nu}\right)$, we approximate the latter by polynomial expansions $f_0^{R,m,z}(x) \approx \sum_{n=0}^{N_\text{poly}} a^{R,m,z}_n x^n$ in the scaling variable $x \equiv v_\mathrm{r} L^{1/\nu}$.
In practice, we use the measured $R_{\mathrm{c}}(V,L)$, $m_{\text{CDW}}^{2}(V,L)$, $Z_{\text{qp}}(V,L)$ as functions of the raw coupling $V$ and the system size $L$ as fitting input, and determine the optimal parameters $V_\mathrm c$, $1/\nu$, $\eta_{\phi}+z$, $\eta_{\psi}$, together with the polynomial coefficients $\{a^{R,m,z}_n\}$ by minimizing the collapse residual~\cite{liu22}. The extracted exponents are found to rapidly stabilize once $N_\text{poly}\ge 4$. We therefore fix $N_\text{poly}=4$ in the data-collapse analysis. 
Figures~\ref{fig:raw_and_rescale_data_t0}(a-c) show the measured data for the correlation ratio $R_\mathrm c$, order parameter $m_\text{CDW}^2$, and quasiparticle weight $Z_\text{qp}$ as a function of $V$ for different system sizes.
The resulting exponents obtained from the scaling-collapse analysis and the corresponding fit qualities $\chi^2$ are given for different minimal lattice sizes $L_\text{min}$ in Table~\ref{tab:collapse}.

%
%

The finite-size exponents exhibit the same convergence trend with increasing $L_{\text{min}}$ and, upon excluding small system sizes, are numerically consistent within statistical uncertainties with those extracted from the crossing-point analysis [Figs.~2(b-d) in the main text], indicating the robustness and mutual consistency of the two independent approaches.
When small system sizes $L \leq 18$ are included, the extracted exponents and the corresponding fitting quality $\chi^{2}$ in the data-collapse analysis exhibit noticeable variations, reflecting the influence of the correction term $L^{-\omega}$. We therefore expect that the crossing-point analysis presented in the main text provides better estimates of the critical exponents in the thermodynamic limit.

As an additional crosscheck, we show in Figs.~\ref{fig:raw_and_rescale_data_t0}(d-f) rescaled observables $R_\mathrm c$, $L^{z+\eta_\phi} m_\text{CDW}^2$, and $L^\eta_\psi Z_\text{qp}$ as functions of $x \equiv (V-V_\mathrm{c})L^{1/\nu}$, using the values of the critical exponents as obtained from the crossing-point analysis presented the main text (see Table~I therein). 
The pronounced deviation for small system sizes highlights the significant finite-size effects.
To illustrate the data collapse for large system sizes, we show in Fig.~\ref{fig:raw_and_rescale_data_t0_2} the same rescaled data, but restricted to those obtained on lattices with system size $L \geq L_\text{min} = 21$.

\section{Charge susceptibility in QBT regime}

In this supplemental section, we discuss the temperature dependence of the uniform charge susceptibility $\chi_\text{uni}(T)$.
For $d \neq z$, a simple scaling argument leads to the scaling form~\cite{hohenadler22}
\begin{align}
\chi_\text{uni}(T) \propto T^{d/z - 1}.
\end{align}
For quadratic band touching (QBT) fermions in two spatial dimensions, we have $d = z =2$. A naive application of the above scaling form suggests a constant charge susceptibility in the low-temperature limit. While this is indeed the correct behavior for noninteracting QBT fermions, the fact that local four-fermion interactions are marginal in the renormalization group sense leads to a logarithmic correction to the above scaling form for interacting QBT fermions.

Consider the single-particle Hamiltonian
\begin{align}
\mathcal H_0(\mathbf k) = \frac{2 k_x k_y \sigma^x + (k_x^2 - k_y^2) \sigma^z}{2 m^*}\,,
\end{align}
describing a single QBT point with effective mass $m^*$ and dispersion relation $\epsilon_{\mathbf k} = \pm \mathbf k^2/(2m^*)$.
It is characterized by a constant density of states $\rho(\epsilon) = m^*/(2\pi)$.
The charge susceptibility may be defined as the response of the particle density $n$ to a change in the chemical potential,
\begin{align}
\chi_\text{uni} = \frac{\partial n(\mu,T)}{\partial \mu}\,.
\end{align}
In the noninteracting limit, the particle density $n$ is given by
\begin{align}
n(\mu,T) = \int_{-\infty}^{\infty} \rmd \epsilon \rho(\epsilon) f(\tfrac{\epsilon - \mu}{k_\mathrm B T})\,,
\end{align}
with $f(x) = (\rme^{x} + 1)^{-1}$ the Fermi-Dirac distribution function, which yields
\begin{align}
\chi_\text{uni} = \frac{m^*}{2\pi}
\end{align}
in the noninteracting limit.

\begin{figure}[tb!]
\includegraphics[width=\linewidth]{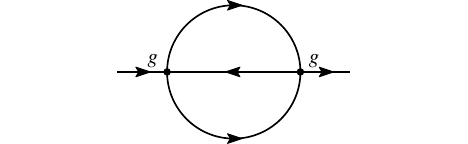}
\caption{Feynman diagram contributing to the fermion self-energy $\Sigma(\mathbf k,\omega)$, renormalizing the effective mass $m^*$ and resulting in a logarithmic temperature dependence of the charge susceptibility in the QBT regime.
\label{fig:selfenergy}
}
\end{figure}

The many-body Hamiltonian in the presence of short-range interactions reads~\cite{sun09}
\begin{align}
H = \int \rmd^2 \mathbf x \left( \Psi^\dagger \mathcal H_0 \Psi + V \psi_1^\dagger \psi_1 \psi_2^\dagger \psi_2 \right),
\end{align}
where $\Psi^\dagger = (\psi^\dagger_1, \psi^\dagger_2)$ is a two-component fermion field and $V$ parametrizes the interaction.
The problem is characterized by a single dimensionless coupling $g = 2 m^* V$, which is marginally relevant in the renormalization group sense for repulsive $g>0$~\cite{sun09,ray20}.
This leads to a logarithmic contribution to the fermion self-energy $\Sigma(\mathbf k,\omega)$, given by the two-loop diagram depicted in Fig.~\ref{fig:selfenergy}. The corresponding loop integration can be carried out explicitly in position space, leading to~\cite{ray18}
\begin{align}
\Sigma(\mathbf k,0) = c g^2 \left[2k_x k_y \sigma^x + (k_x^2 - k_y^2) \sigma^z \right] \ln \frac{\Lambda}{\lambda} + \mathcal O(k^3)\,,
\end{align}
where $c$ corresponds to a nonuniversal constant and $\Lambda$ ($\lambda$) the ultraviolet (infrared) cutoff.
The above form of the self-energy implies a logarithmic contribution to the mass renormalization
\begin{align}
m^{*\prime} = m^* \left[1 - c g^2 \ln \frac{\Lambda}{\lambda} + \mathcal O(g^3)\right].
\end{align}
At finite temperatures, the infrared cutoff is controlled by the thermal scale, $\lambda \propto T^{1/z}$. Consequently, the charge susceptibility acquires a logarithmic temperature dependence,
\begin{align}
\chi_\text{uni}(T) = \frac{m^*(T)}{2\pi} = \frac{1}{2\pi} \left[ m_0^* + c z g^2 \ln \frac{T}{T_0} + \mathcal O(g^3) \right],
\end{align}
in agreement with the Monte Carlo results at elevated temperatures $T > T_\text{cross}$, presented in the main text.

\section{Charge susceptibility: Additional data}

In this supplemental section, we present additional data of the finite-temperature uniform charge susceptibility $\chi_{\text{uni}}(T)$ to support the results discussed in the main text. 

\begin{figure}[t!]
\includegraphics[width=\linewidth]{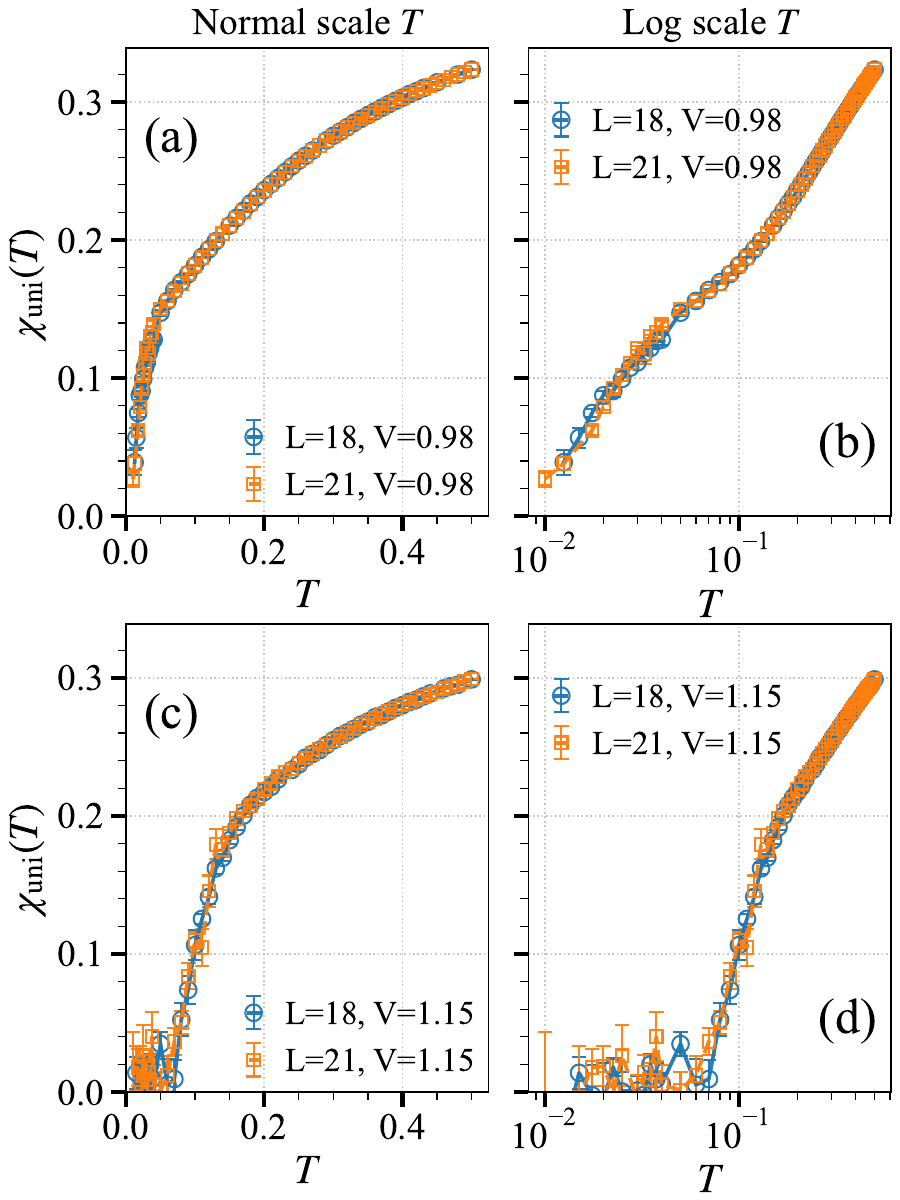}
\caption{%
(a)~Uniform charge susceptibility $\chi_{\text{uni}}$ as a function of temperature $T$ at $V = 0.98$ and two different system sizes $L$, shown on a linear temperature scale.
(b)~Same as (a), but on a logarithmic temperature scale.
(c)~Same as (a), but for $V=1.15$.
(d)~Same as (b), but for $V=1.15$.
For both values of $V$ shown, the curves for system sizes $L = 18$ and $L = 21$ are essentially indistinguishable, indicating negligible finite-size effects at the temperatures considered.
%
\label{fig:size_effect}}
\end{figure}

In Fig.~\ref{fig:size_effect}, we examine the finite-size convergence of $\chi_{\text{uni}}(T)$ on both linear and logarithmic temperature scales. The data show that, within the parameter range $V \in [0.94, 1.15]$ and for the temperatures considered, the uniform charge susceptibility exhibits excellent convergence for system sizes $L \ge 18$. This indicates that the results presented in the main text for $L=21$ are effectively free from finite-size effects for the purposes of our analysis.

\begin{figure}[t!]
\includegraphics[width=\linewidth]{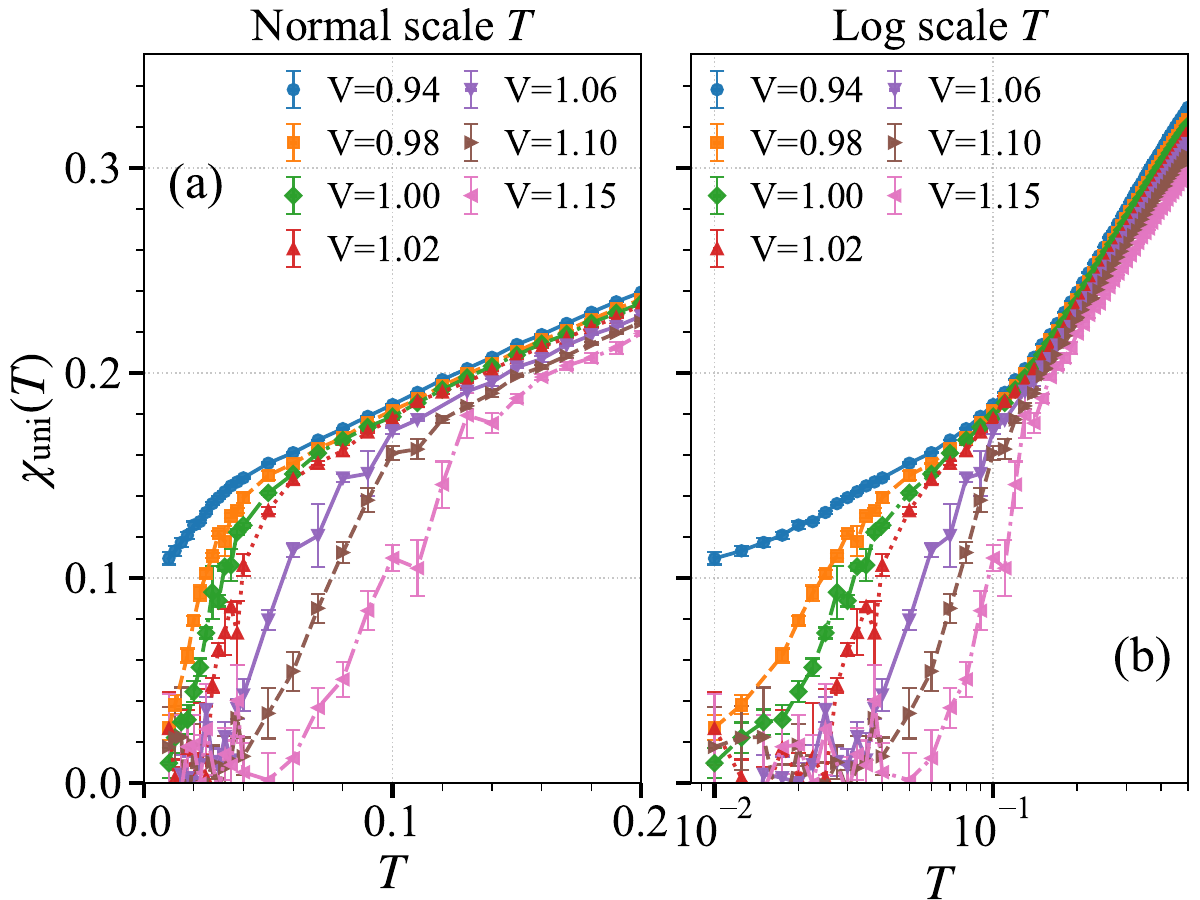}
\caption{%
(a)~Uniform charge susceptibility $\chi_{\text{uni}}$ as a function of temperature $T$ for $L=21$ and different $V$, shown on a linear temperature scale.
(b)~Same as (a), but on a logarithmic temperature scale.
%
%
\label{fig:uni_chi_ft_total}}
\end{figure}

Figures~\ref{fig:uni_chi_ft_total}(a) and \ref{fig:uni_chi_ft_total}(b) show $\chi_{\text{uni}}$ as a function of $T$ for different fixed interaction strengths $V$ on linear and logarithmic temperature scales, respectively. At high temperatures, $\chi_{\text{uni}}$ shows a logarithmic temperature dependence, consistent with the expected behavior for QBT fermions. At very low temperatures, $\chi_{\text{uni}}(T)$ vanishes due to the gap opening in the CDW ordered phase. In the intermediate regime, $\chi_{\text{uni}}(T)$ exhibits a linear temperature dependence, reflecting the emergent relativistic symmetric of the interaction-driven Dirac semimetal phase. 

\begin{figure}[b!]
\includegraphics[width=\linewidth]{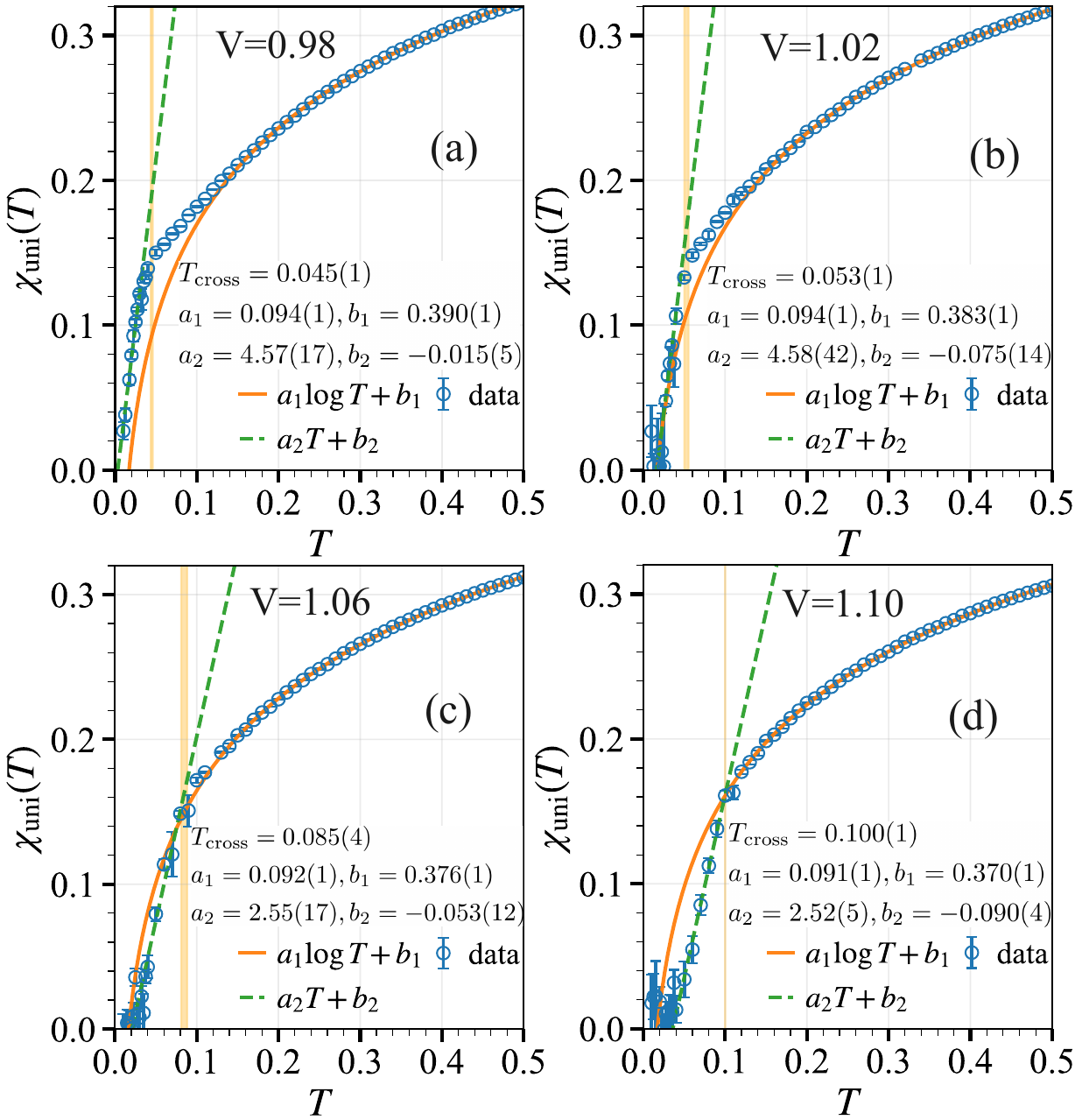}
\caption{%
(a)~Uniform charge susceptibility $\chi_{\text{uni}}$ as a function of $T$ for $V=0.98$ and $L=21$.  The shaded yellow region indicates the confidence interval of the crossover temperature $T_{\text{cross}}$, determined by the equal-residual criterion.
(b)~Same as (a), but for $V=1.02$.
(c)~Same as (a), but for $V=1.06$.
(d)~Same as (a), but for $V=1.10$.
}
\label{fig:t_cross_V}
\end{figure}

Finally, we provide supplemental analysis of $\chi_{\text{uni}}$ for different fixed interaction strengths $V$, shown in Fig.~\ref{fig:t_cross_V}. The low-temperature data are fitted using a linear form $a_1 T + b_1$, while the high-temperature data follow a logarithmic dependence $a_2 \log T + b_2$. The crossover temperature separating the DSM and QBT regimes is determined via the equal-residual criterion. The corresponding fits and residual analysis are presented for completeness.

\FloatBarrier

%% file: qbt2d-honeycomb-qmc.bbl
\begin{thebibliography}{42}%
\makeatletter
\providecommand \@ifxundefined [1]{%
 \@ifx{#1\undefined}
}%
\providecommand \@ifnum [1]{%
 \ifnum #1\expandafter \@firstoftwo
 \else \expandafter \@secondoftwo
 \fi
}%
\providecommand \@ifx [1]{%
 \ifx #1\expandafter \@firstoftwo
 \else \expandafter \@secondoftwo
 \fi
}%
\providecommand \natexlab [1]{#1}%
\providecommand \enquote  [1]{``#1''}%
\providecommand \bibnamefont  [1]{#1}%
\providecommand \bibfnamefont [1]{#1}%
\providecommand \citenamefont [1]{#1}%
\providecommand \href@noop [0]{\@secondoftwo}%
\providecommand \href [0]{\begingroup \@sanitize@url \@href}%
\providecommand \@href[1]{\@@startlink{#1}\@@href}%
\providecommand \@@href[1]{\endgroup#1\@@endlink}%
\providecommand \@sanitize@url [0]{\catcode `\\12\catcode `\$12\catcode
  `\&12\catcode `\#12\catcode `\^12\catcode `\_12\catcode `\%12\relax}%
\providecommand \@@startlink[1]{}%
\providecommand \@@endlink[0]{}%
\providecommand \url  [0]{\begingroup\@sanitize@url \@url }%
\providecommand \@url [1]{\endgroup\@href {#1}{\urlprefix }}%
\providecommand \urlprefix  [0]{URL }%
\providecommand \Eprint [0]{\href }%
\providecommand \doibase [0]{https://doi.org/}%
\providecommand \selectlanguage [0]{\@gobble}%
\providecommand \bibinfo  [0]{\@secondoftwo}%
\providecommand \bibfield  [0]{\@secondoftwo}%
\providecommand \translation [1]{[#1]}%
\providecommand \BibitemOpen [0]{}%
\providecommand \bibitemStop [0]{}%
\providecommand \bibitemNoStop [0]{.\EOS\space}%
\providecommand \EOS [0]{\spacefactor3000\relax}%
\providecommand \BibitemShut  [1]{\csname bibitem#1\endcsname}%
\let\auto@bib@innerbib\@empty
\bibitem [{\citenamefont {Sachdev}(2011)}]{sachdevbook}%
  \BibitemOpen
  \bibfield  {author} {\bibinfo {author} {\bibfnamefont {S.}~\bibnamefont
  {Sachdev}},\ }\href {https://doi.org/10.1017/CBO9780511973765} {\emph
  {\bibinfo {title} {Quantum Phase Transitions}}},\ \bibinfo {edition} {2nd}\
  ed.\ (\bibinfo  {publisher} {Cambridge University Press},\ \bibinfo {address}
  {Cambridge, UK},\ \bibinfo {year} {2011})\BibitemShut {NoStop}%
\bibitem [{\citenamefont {Pelissetto}\ and\ \citenamefont
  {Vicari}(2002)}]{pelissetto02}%
  \BibitemOpen
  \bibfield  {author} {\bibinfo {author} {\bibfnamefont {A.}~\bibnamefont
  {Pelissetto}}\ and\ \bibinfo {author} {\bibfnamefont {E.}~\bibnamefont
  {Vicari}},\ }\bibfield  {title} {\bibinfo {title} {Critical phenomena and
  renormalization-group theory},\ }\href
  {https://doi.org/https://doi.org/10.1016/S0370-1573(02)00219-3} {\bibfield
  {journal} {\bibinfo  {journal} {Phys. Rep.}\ }\textbf {\bibinfo {volume}
  {368}},\ \bibinfo {pages} {549} (\bibinfo {year} {2002})}\BibitemShut
  {NoStop}%
\bibitem [{\citenamefont {Lou}\ \emph {et~al.}(2007)\citenamefont {Lou},
  \citenamefont {Sandvik},\ and\ \citenamefont {Balents}}]{lou07}%
  \BibitemOpen
  \bibfield  {author} {\bibinfo {author} {\bibfnamefont {J.}~\bibnamefont
  {Lou}}, \bibinfo {author} {\bibfnamefont {A.~W.}\ \bibnamefont {Sandvik}},\
  and\ \bibinfo {author} {\bibfnamefont {L.}~\bibnamefont {Balents}},\
  }\bibfield  {title} {\bibinfo {title} {Emergence of U(1) Symmetry in the 3D
  $XY$ Model with ${Z}_{q}$ Anisotropy},\ }\href
  {https://doi.org/10.1103/PhysRevLett.99.207203} {\bibfield  {journal}
  {\bibinfo  {journal} {Phys. Rev. Lett.}\ }\textbf {\bibinfo {volume} {99}},\
  \bibinfo {pages} {207203} (\bibinfo {year} {2007})}\BibitemShut {NoStop}%
\bibitem [{\citenamefont {Patil}\ \emph {et~al.}(2021)\citenamefont {Patil},
  \citenamefont {Shao},\ and\ \citenamefont {Sandvik}}]{patil21}%
  \BibitemOpen
  \bibfield  {author} {\bibinfo {author} {\bibfnamefont {P.}~\bibnamefont
  {Patil}}, \bibinfo {author} {\bibfnamefont {H.}~\bibnamefont {Shao}},\ and\
  \bibinfo {author} {\bibfnamefont {A.~W.}\ \bibnamefont {Sandvik}},\
  }\bibfield  {title} {\bibinfo {title} {Unconventional U(1) to ${Z}_{q}$
  crossover in quantum and classical $q$-state clock models},\ }\href
  {https://doi.org/10.1103/PhysRevB.103.054418} {\bibfield  {journal} {\bibinfo
   {journal} {Phys. Rev. B}\ }\textbf {\bibinfo {volume} {103}},\ \bibinfo
  {pages} {054418} (\bibinfo {year} {2021})}\BibitemShut {NoStop}%
\bibitem [{\citenamefont {Senthil}(2023)}]{senthil23}%
  \BibitemOpen
  \bibfield  {author} {\bibinfo {author} {\bibfnamefont {T.}~\bibnamefont
  {Senthil}},\ }\bibinfo {title} {Deconfined Quantum Critical Points: A
  Review},\ in\ \href {https://doi.org/doi:10.1142/9789811282386_0014} {\emph
  {\bibinfo {booktitle} {50 Years of the Renormalization Group}}}\ (\bibinfo
  {publisher} {World Scientific},\ \bibinfo {year} {2023})\ pp.\ \bibinfo
  {pages} {169--195},\ \Eprint {https://arxiv.org/abs/2306.12638}
  {arXiv:2306.12638}\BibitemShut {NoStop}%
\bibitem [{\citenamefont {Roy}\ \emph {et~al.}(2016)\citenamefont {Roy},
  \citenamefont {Juri{\v c}i{\'c}},\ and\ \citenamefont {Herbut}}]{roy16}%
  \BibitemOpen
  \bibfield  {author} {\bibinfo {author} {\bibfnamefont {B.}~\bibnamefont
  {Roy}}, \bibinfo {author} {\bibfnamefont {V.}~\bibnamefont {Juri{\v
  c}i{\'c}}},\ and\ \bibinfo {author} {\bibfnamefont {I.~F.}\ \bibnamefont
  {Herbut}},\ }\bibfield  {title} {\bibinfo {title} {Emergent Lorentz symmetry
  near fermionic quantum critical points in two and three dimensions},\ }\href
  {https://doi.org/10.1007/JHEP04(2016)018} {\bibfield  {journal} {\bibinfo
  {journal} {J. High Energy Phys.}\ }4\bibfield  {volume} {\bibinfo  {volume} {
  (2016)}\ }18}\BibitemShut {NoStop}%
\bibitem [{\citenamefont {Biedermann}\ and\ \citenamefont
  {Janssen}(2026)}]{biedermann26}%
  \BibitemOpen
  \bibfield  {author} {\bibinfo {author} {\bibfnamefont {J.}~\bibnamefont
  {Biedermann}}\ and\ \bibinfo {author} {\bibfnamefont {L.}~\bibnamefont
  {Janssen}},\ }\bibinfo {title} {Dirac quantum criticality in twisted double
  bilayer transition metal dichalcogenides},\ \Eprint
  {https://arxiv.org/abs/2509.04561} {arXiv:2509.04561}\BibitemShut {NoStop}%
\bibitem [{\citenamefont {Lee}(2007)}]{lee07}%
  \BibitemOpen
  \bibfield  {author} {\bibinfo {author} {\bibfnamefont {S.-S.}\ \bibnamefont
  {Lee}},\ }\bibfield  {title} {\bibinfo {title} {Emergence of supersymmetry at
  a critical point of a lattice model},\ }\href
  {https://doi.org/10.1103/PhysRevB.76.075103} {\bibfield  {journal} {\bibinfo
  {journal} {Phys. Rev. B}\ }\textbf {\bibinfo {volume} {76}},\ \bibinfo
  {pages} {075103} (\bibinfo {year} {2007})}\BibitemShut {NoStop}%
\bibitem [{\citenamefont {Grover}\ \emph {et~al.}(2014)\citenamefont {Grover},
  \citenamefont {Sheng},\ and\ \citenamefont {Vishwanath}}]{grover14}%
  \BibitemOpen
  \bibfield  {author} {\bibinfo {author} {\bibfnamefont {T.}~\bibnamefont
  {Grover}}, \bibinfo {author} {\bibfnamefont {D.~N.}\ \bibnamefont {Sheng}},\
  and\ \bibinfo {author} {\bibfnamefont {A.}~\bibnamefont {Vishwanath}},\
  }\bibfield  {title} {\bibinfo {title} {Emergent Space-Time Supersymmetry at
  the Boundary of a Topological Phase},\ }\href
  {https://doi.org/10.1126/science.1248253} {\bibfield  {journal} {\bibinfo
  {journal} {Science}\ }\textbf {\bibinfo {volume} {344}},\ \bibinfo {pages}
  {280} (\bibinfo {year} {2014})}\BibitemShut {NoStop}%
\bibitem [{\citenamefont {Schuler}\ \emph {et~al.}(2023)\citenamefont
  {Schuler}, \citenamefont {Henry}, \citenamefont {Lu},\ and\ \citenamefont
  {Läuchli}}]{schuler23}%
  \BibitemOpen
  \bibfield  {author} {\bibinfo {author} {\bibfnamefont {M.}~\bibnamefont
  {Schuler}}, \bibinfo {author} {\bibfnamefont {L.-P.}\ \bibnamefont {Henry}},
  \bibinfo {author} {\bibfnamefont {Y.-M.}\ \bibnamefont {Lu}},\ and\ \bibinfo
  {author} {\bibfnamefont {A.~M.}\ \bibnamefont {Läuchli}},\ }\bibfield
  {title} {\bibinfo {title} {{Emergent XY* transition driven by symmetry
  fractionalization and anyon condensation}},\ }\href
  {https://doi.org/10.21468/SciPostPhys.14.1.001} {\bibfield  {journal}
  {\bibinfo  {journal} {SciPost Phys.}\ }\textbf {\bibinfo {volume} {14}},\
  \bibinfo {pages} {001} (\bibinfo {year} {2023})}\BibitemShut {NoStop}%
\bibitem [{\citenamefont {Nahum}\ \emph {et~al.}(2015)\citenamefont {Nahum},
  \citenamefont {Serna}, \citenamefont {Chalker}, \citenamefont {Ortu\~no},\
  and\ \citenamefont {Somoza}}]{nahum15b}%
  \BibitemOpen
  \bibfield  {author} {\bibinfo {author} {\bibfnamefont {A.}~\bibnamefont
  {Nahum}}, \bibinfo {author} {\bibfnamefont {P.}~\bibnamefont {Serna}},
  \bibinfo {author} {\bibfnamefont {J.~T.}\ \bibnamefont {Chalker}}, \bibinfo
  {author} {\bibfnamefont {M.}~\bibnamefont {Ortu\~no}},\ and\ \bibinfo
  {author} {\bibfnamefont {A.~M.}\ \bibnamefont {Somoza}},\ }\bibfield  {title}
  {\bibinfo {title} {Emergent SO(5) Symmetry at the N\'eel to
  Valence-Bond-Solid Transition},\ }\href
  {https://doi.org/10.1103/PhysRevLett.115.267203} {\bibfield  {journal}
  {\bibinfo  {journal} {Phys. Rev. Lett.}\ }\textbf {\bibinfo {volume} {115}},\
  \bibinfo {pages} {267203} (\bibinfo {year} {2015})}\BibitemShut {NoStop}%
\bibitem [{\citenamefont {Nakayama}\ and\ \citenamefont
  {Ohtsuki}(2016)}]{nakayama16}%
  \BibitemOpen
  \bibfield  {author} {\bibinfo {author} {\bibfnamefont {Y.}~\bibnamefont
  {Nakayama}}\ and\ \bibinfo {author} {\bibfnamefont {T.}~\bibnamefont
  {Ohtsuki}},\ }\bibfield  {title} {\bibinfo {title} {Necessary Condition for
  Emergent Symmetry from the Conformal Bootstrap},\ }\href
  {https://doi.org/10.1103/PhysRevLett.117.131601} {\bibfield  {journal}
  {\bibinfo  {journal} {Phys. Rev. Lett.}\ }\textbf {\bibinfo {volume} {117}},\
  \bibinfo {pages} {131601} (\bibinfo {year} {2016})}\BibitemShut {NoStop}%
\bibitem [{\citenamefont {Poland}\ \emph {et~al.}(2019)\citenamefont {Poland},
  \citenamefont {Rychkov},\ and\ \citenamefont {Vichi}}]{poland19}%
  \BibitemOpen
  \bibfield  {author} {\bibinfo {author} {\bibfnamefont {D.}~\bibnamefont
  {Poland}}, \bibinfo {author} {\bibfnamefont {S.}~\bibnamefont {Rychkov}},\
  and\ \bibinfo {author} {\bibfnamefont {A.}~\bibnamefont {Vichi}},\ }\bibfield
   {title} {\bibinfo {title} {The conformal bootstrap: Theory, numerical
  techniques, and applications},\ }\href
  {https://doi.org/10.1103/RevModPhys.91.015002} {\bibfield  {journal}
  {\bibinfo  {journal} {Rev. Mod. Phys.}\ }\textbf {\bibinfo {volume} {91}},\
  \bibinfo {pages} {015002} (\bibinfo {year} {2019})}\BibitemShut {NoStop}%
\bibitem [{\citenamefont {Assaad}\ and\ \citenamefont
  {Herbut}(2013)}]{assaad13}%
  \BibitemOpen
  \bibfield  {author} {\bibinfo {author} {\bibfnamefont {F.~F.}\ \bibnamefont
  {Assaad}}\ and\ \bibinfo {author} {\bibfnamefont {I.~F.}\ \bibnamefont
  {Herbut}},\ }\bibfield  {title} {\bibinfo {title} {Pinning the Order: The
  Nature of Quantum Criticality in the Hubbard Model on Honeycomb Lattice},\
  }\href {https://doi.org/10.1103/PhysRevX.3.031010} {\bibfield  {journal}
  {\bibinfo  {journal} {Phys. Rev. X}\ }\textbf {\bibinfo {volume} {3}},\
  \bibinfo {pages} {031010} (\bibinfo {year} {2013})}\BibitemShut {NoStop}%
\bibitem [{\citenamefont {Parisen~Toldin}\ \emph {et~al.}(2015)\citenamefont
  {Parisen~Toldin}, \citenamefont {Hohenadler}, \citenamefont {Assaad},\ and\
  \citenamefont {Herbut}}]{toldin15}%
  \BibitemOpen
  \bibfield  {author} {\bibinfo {author} {\bibfnamefont {F.}~\bibnamefont
  {Parisen~Toldin}}, \bibinfo {author} {\bibfnamefont {M.}~\bibnamefont
  {Hohenadler}}, \bibinfo {author} {\bibfnamefont {F.~F.}\ \bibnamefont
  {Assaad}},\ and\ \bibinfo {author} {\bibfnamefont {I.~F.}\ \bibnamefont
  {Herbut}},\ }\bibfield  {title} {\bibinfo {title} {Fermionic quantum
  criticality in honeycomb and $\ensuremath{\pi}$-flux Hubbard models:
  Finite-size scaling of renormalization-group-invariant observables from
  quantum Monte Carlo},\ }\href {https://doi.org/10.1103/PhysRevB.91.165108}
  {\bibfield  {journal} {\bibinfo  {journal} {Phys. Rev. B}\ }\textbf {\bibinfo
  {volume} {91}},\ \bibinfo {pages} {165108} (\bibinfo {year}
  {2015})}\BibitemShut {NoStop}%
\bibitem [{\citenamefont {Otsuka}\ \emph {et~al.}(2016)\citenamefont {Otsuka},
  \citenamefont {Yunoki},\ and\ \citenamefont {Sorella}}]{otsuka16}%
  \BibitemOpen
  \bibfield  {author} {\bibinfo {author} {\bibfnamefont {Y.}~\bibnamefont
  {Otsuka}}, \bibinfo {author} {\bibfnamefont {S.}~\bibnamefont {Yunoki}},\
  and\ \bibinfo {author} {\bibfnamefont {S.}~\bibnamefont {Sorella}},\
  }\bibfield  {title} {\bibinfo {title} {Universal Quantum Criticality in the
  Metal-Insulator Transition of Two-Dimensional Interacting Dirac Electrons},\
  }\href {https://doi.org/10.1103/PhysRevX.6.011029} {\bibfield  {journal}
  {\bibinfo  {journal} {Phys. Rev. X}\ }\textbf {\bibinfo {volume} {6}},\
  \bibinfo {pages} {011029} (\bibinfo {year} {2016})}\BibitemShut {NoStop}%
\bibitem [{\citenamefont {Herbut}(2006)}]{herbut06}%
  \BibitemOpen
  \bibfield  {author} {\bibinfo {author} {\bibfnamefont {I.~F.}\ \bibnamefont
  {Herbut}},\ }\bibfield  {title} {\bibinfo {title} {Interactions and Phase
  Transitions on Graphene's Honeycomb Lattice},\ }\href
  {https://doi.org/10.1103/PhysRevLett.97.146401} {\bibfield  {journal}
  {\bibinfo  {journal} {Phys. Rev. Lett.}\ }\textbf {\bibinfo {volume} {97}},\
  \bibinfo {pages} {146401} (\bibinfo {year} {2006})}\BibitemShut {NoStop}%
\bibitem [{\citenamefont {Herbut}\ \emph {et~al.}(2009)\citenamefont {Herbut},
  \citenamefont {Juri\ifmmode \check{c}\else \v{c}\fi{}i\ifmmode~\acute{c}\else
  \'{c}\fi{}},\ and\ \citenamefont {Vafek}}]{herbut09}%
  \BibitemOpen
  \bibfield  {author} {\bibinfo {author} {\bibfnamefont {I.~F.}\ \bibnamefont
  {Herbut}}, \bibinfo {author} {\bibfnamefont {V.}~\bibnamefont {Juri\ifmmode
  \check{c}\else \v{c}\fi{}i\ifmmode~\acute{c}\else \'{c}\fi{}}},\ and\
  \bibinfo {author} {\bibfnamefont {O.}~\bibnamefont {Vafek}},\ }\bibfield
  {title} {\bibinfo {title} {Relativistic Mott criticality in graphene},\
  }\href {https://doi.org/10.1103/PhysRevB.80.075432} {\bibfield  {journal}
  {\bibinfo  {journal} {Phys. Rev. B}\ }\textbf {\bibinfo {volume} {80}},\
  \bibinfo {pages} {075432} (\bibinfo {year} {2009})}\BibitemShut {NoStop}%
\bibitem [{\citenamefont {Janssen}\ and\ \citenamefont
  {Herbut}(2014)}]{janssen14}%
  \BibitemOpen
  \bibfield  {author} {\bibinfo {author} {\bibfnamefont {L.}~\bibnamefont
  {Janssen}}\ and\ \bibinfo {author} {\bibfnamefont {I.~F.}\ \bibnamefont
  {Herbut}},\ }\bibfield  {title} {\bibinfo {title} {Antiferromagnetic critical
  point on graphene's honeycomb lattice: A functional renormalization group
  approach},\ }\href {https://doi.org/10.1103/PhysRevB.89.205403} {\bibfield
  {journal} {\bibinfo  {journal} {Phys. Rev. B}\ }\textbf {\bibinfo {volume}
  {89}},\ \bibinfo {pages} {205403} (\bibinfo {year} {2014})}\BibitemShut
  {NoStop}%
\bibitem [{\citenamefont {Ladovrechis}\ \emph {et~al.}(2023)\citenamefont
  {Ladovrechis}, \citenamefont {Ray}, \citenamefont {Meng},\ and\ \citenamefont
  {Janssen}}]{ladovrechis23}%
  \BibitemOpen
  \bibfield  {author} {\bibinfo {author} {\bibfnamefont {K.}~\bibnamefont
  {Ladovrechis}}, \bibinfo {author} {\bibfnamefont {S.}~\bibnamefont {Ray}},
  \bibinfo {author} {\bibfnamefont {T.}~\bibnamefont {Meng}},\ and\ \bibinfo
  {author} {\bibfnamefont {L.}~\bibnamefont {Janssen}},\ }\bibfield  {title}
  {\bibinfo {title} {Gross-Neveu-Heisenberg criticality from
  $2+\ensuremath{\epsilon}$ expansion},\ }\href
  {https://doi.org/10.1103/PhysRevB.107.035151} {\bibfield  {journal} {\bibinfo
   {journal} {Phys. Rev. B}\ }\textbf {\bibinfo {volume} {107}},\ \bibinfo
  {pages} {035151} (\bibinfo {year} {2023})}\BibitemShut {NoStop}%
\bibitem [{\citenamefont {Lang}\ and\ \citenamefont
  {L\"auchli}(2025)}]{lang25}%
  \BibitemOpen
  \bibfield  {author} {\bibinfo {author} {\bibfnamefont {T.~C.}\ \bibnamefont
  {Lang}}\ and\ \bibinfo {author} {\bibfnamefont {A.~M.}\ \bibnamefont
  {L\"auchli}},\ }\bibfield  {title} {\bibinfo {title} {Chiral Heisenberg
  Gross-Neveu-Yukawa criticality: Honeycomb versus SLAC fermions},\ }\href
  {https://doi.org/10.1103/vlgd-7ln8} {\bibfield  {journal} {\bibinfo
  {journal} {Phys. Rev. B}\ }\textbf {\bibinfo {volume} {112}},\ \bibinfo
  {pages} {245121} (\bibinfo {year} {2025})}\BibitemShut {NoStop}%
\bibitem [{\citenamefont {Wang}\ \emph {et~al.}(2026)\citenamefont {Wang},
  \citenamefont {Sun}, \citenamefont {He},\ and\ \citenamefont {Xu}}]{wang26}%
  \BibitemOpen
  \bibfield  {author} {\bibinfo {author} {\bibfnamefont {F.-H.}\ \bibnamefont
  {Wang}}, \bibinfo {author} {\bibfnamefont {F.}~\bibnamefont {Sun}}, \bibinfo
  {author} {\bibfnamefont {C.}~\bibnamefont {He}},\ and\ \bibinfo {author}
  {\bibfnamefont {X.~Y.}\ \bibnamefont {Xu}},\ }\bibinfo {title} {Resolving
  Quantum Criticality in the Honeycomb Hubbard Model},\ \Eprint
  {https://arxiv.org/abs/2602.03656} {arXiv:2602.03656}\BibitemShut {NoStop}%
\bibitem [{\citenamefont {Vafek}(2010)}]{vafek10}%
  \BibitemOpen
  \bibfield  {author} {\bibinfo {author} {\bibfnamefont {O.}~\bibnamefont
  {Vafek}},\ }\bibfield  {title} {\bibinfo {title} {Interacting fermions on the
  honeycomb bilayer: From weak to strong coupling},\ }\href
  {https://doi.org/10.1103/PhysRevB.82.205106} {\bibfield  {journal} {\bibinfo
  {journal} {Phys. Rev. B}\ }\textbf {\bibinfo {volume} {82}},\ \bibinfo
  {pages} {205106} (\bibinfo {year} {2010})}\BibitemShut {NoStop}%
\bibitem [{\citenamefont {Ray}\ \emph {et~al.}(2018)\citenamefont {Ray},
  \citenamefont {Vojta},\ and\ \citenamefont {Janssen}}]{ray18}%
  \BibitemOpen
  \bibfield  {author} {\bibinfo {author} {\bibfnamefont {S.}~\bibnamefont
  {Ray}}, \bibinfo {author} {\bibfnamefont {M.}~\bibnamefont {Vojta}},\ and\
  \bibinfo {author} {\bibfnamefont {L.}~\bibnamefont {Janssen}},\ }\bibfield
  {title} {\bibinfo {title} {Quantum critical behavior of two-dimensional Fermi
  systems with quadratic band touching},\ }\href
  {https://doi.org/10.1103/PhysRevB.98.245128} {\bibfield  {journal} {\bibinfo
  {journal} {Phys. Rev. B}\ }\textbf {\bibinfo {volume} {98}},\ \bibinfo
  {pages} {245128} (\bibinfo {year} {2018})}\BibitemShut {NoStop}%
\bibitem [{\citenamefont {Pujari}\ \emph {et~al.}(2016)\citenamefont {Pujari},
  \citenamefont {Lang}, \citenamefont {Murthy},\ and\ \citenamefont
  {Kaul}}]{pujari16}%
  \BibitemOpen
  \bibfield  {author} {\bibinfo {author} {\bibfnamefont {S.}~\bibnamefont
  {Pujari}}, \bibinfo {author} {\bibfnamefont {T.~C.}\ \bibnamefont {Lang}},
  \bibinfo {author} {\bibfnamefont {G.}~\bibnamefont {Murthy}},\ and\ \bibinfo
  {author} {\bibfnamefont {R.~K.}\ \bibnamefont {Kaul}},\ }\bibfield  {title}
  {\bibinfo {title} {Interaction-Induced Dirac Fermions from Quadratic Band
  Touching in Bilayer Graphene},\ }\href
  {https://doi.org/10.1103/PhysRevLett.117.086404} {\bibfield  {journal}
  {\bibinfo  {journal} {Phys. Rev. Lett.}\ }\textbf {\bibinfo {volume} {117}},\
  \bibinfo {pages} {086404} (\bibinfo {year} {2016})}\BibitemShut {NoStop}%
\bibitem [{\citenamefont {Wei}\ \emph {et~al.}(2016)\citenamefont {Wei},
  \citenamefont {Wu}, \citenamefont {Li}, \citenamefont {Zhang},\ and\
  \citenamefont {Xiang}}]{wei16}%
  \BibitemOpen
  \bibfield  {author} {\bibinfo {author} {\bibfnamefont {Z.~C.}\ \bibnamefont
  {Wei}}, \bibinfo {author} {\bibfnamefont {C.}~\bibnamefont {Wu}}, \bibinfo
  {author} {\bibfnamefont {Y.}~\bibnamefont {Li}}, \bibinfo {author}
  {\bibfnamefont {S.}~\bibnamefont {Zhang}},\ and\ \bibinfo {author}
  {\bibfnamefont {T.}~\bibnamefont {Xiang}},\ }\bibfield  {title} {\bibinfo
  {title} {Majorana Positivity and the Fermion Sign Problem of Quantum Monte
  Carlo Simulations},\ }\href {https://doi.org/10.1103/PhysRevLett.116.250601}
  {\bibfield  {journal} {\bibinfo  {journal} {Phys. Rev. Lett.}\ }\textbf
  {\bibinfo {volume} {116}},\ \bibinfo {pages} {250601} (\bibinfo {year}
  {2016})}\BibitemShut {NoStop}%
\bibitem [{\citenamefont {Sun}\ and\ \citenamefont {Xu}(2025)}]{sun25}%
  \BibitemOpen
  \bibfield  {author} {\bibinfo {author} {\bibfnamefont {F.}~\bibnamefont
  {Sun}}\ and\ \bibinfo {author} {\bibfnamefont {X.~Y.}\ \bibnamefont {Xu}},\
  }\bibfield  {title} {\bibinfo {title} {{Boosting determinant quantum Monte
  Carlo with submatrix updates: Unveiling the phase diagram of the 3D Hubbard
  model}},\ }\href {https://doi.org/10.21468/SciPostPhys.18.2.055} {\bibfield
  {journal} {\bibinfo  {journal} {SciPost Phys.}\ }\textbf {\bibinfo {volume}
  {18}},\ \bibinfo {pages} {055} (\bibinfo {year} {2025})}\BibitemShut
  {NoStop}%
\bibitem [{\citenamefont {Liu}\ \emph {et~al.}(2022)\citenamefont {Liu},
  \citenamefont {Vojta}, \citenamefont {Assaad},\ and\ \citenamefont
  {Janssen}}]{liu22}%
  \BibitemOpen
  \bibfield  {author} {\bibinfo {author} {\bibfnamefont {Z.~H.}\ \bibnamefont
  {Liu}}, \bibinfo {author} {\bibfnamefont {M.}~\bibnamefont {Vojta}}, \bibinfo
  {author} {\bibfnamefont {F.~F.}\ \bibnamefont {Assaad}},\ and\ \bibinfo
  {author} {\bibfnamefont {L.}~\bibnamefont {Janssen}},\ }\bibfield  {title}
  {\bibinfo {title} {Metallic and Deconfined Quantum Criticality in Dirac
  Systems},\ }\href {https://doi.org/10.1103/PhysRevLett.128.087201} {\bibfield
   {journal} {\bibinfo  {journal} {Phys. Rev. Lett.}\ }\textbf {\bibinfo
  {volume} {128}},\ \bibinfo {pages} {087201} (\bibinfo {year}
  {2022})}\BibitemShut {NoStop}%
\bibitem [{\citenamefont {Liu}\ \emph {et~al.}(2024)\citenamefont {Liu},
  \citenamefont {Vojta}, \citenamefont {Assaad},\ and\ \citenamefont
  {Janssen}}]{liu24}%
  \BibitemOpen
  \bibfield  {author} {\bibinfo {author} {\bibfnamefont {Z.~H.}\ \bibnamefont
  {Liu}}, \bibinfo {author} {\bibfnamefont {M.}~\bibnamefont {Vojta}}, \bibinfo
  {author} {\bibfnamefont {F.~F.}\ \bibnamefont {Assaad}},\ and\ \bibinfo
  {author} {\bibfnamefont {L.}~\bibnamefont {Janssen}},\ }\bibfield  {title}
  {\bibinfo {title} {Critical properties of metallic and deconfined quantum
  phase transitions in Dirac systems},\ }\href
  {https://doi.org/10.1103/PhysRevB.110.125123} {\bibfield  {journal} {\bibinfo
   {journal} {Phys. Rev. B}\ }\textbf {\bibinfo {volume} {110}},\ \bibinfo
  {pages} {125123} (\bibinfo {year} {2024})}\BibitemShut {NoStop}%
\bibitem [{\citenamefont {Campostrini}\ \emph {et~al.}(2014)\citenamefont
  {Campostrini}, \citenamefont {Pelissetto},\ and\ \citenamefont
  {Vicari}}]{campostrini14}%
  \BibitemOpen
  \bibfield  {author} {\bibinfo {author} {\bibfnamefont {M.}~\bibnamefont
  {Campostrini}}, \bibinfo {author} {\bibfnamefont {A.}~\bibnamefont
  {Pelissetto}},\ and\ \bibinfo {author} {\bibfnamefont {E.}~\bibnamefont
  {Vicari}},\ }\bibfield  {title} {\bibinfo {title} {Finite-size scaling at
  quantum transitions},\ }\href {https://doi.org/10.1103/PhysRevB.89.094516}
  {\bibfield  {journal} {\bibinfo  {journal} {Phys. Rev. B}\ }\textbf {\bibinfo
  {volume} {89}},\ \bibinfo {pages} {094516} (\bibinfo {year}
  {2014})}\BibitemShut {NoStop}%
\bibitem [{\citenamefont {Harada}(2011)}]{harada11}%
  \BibitemOpen
  \bibfield  {author} {\bibinfo {author} {\bibfnamefont {K.}~\bibnamefont
  {Harada}},\ }\bibfield  {title} {\bibinfo {title} {Bayesian inference in the
  scaling analysis of critical phenomena},\ }\href
  {https://doi.org/10.1103/PhysRevE.84.056704} {\bibfield  {journal} {\bibinfo
  {journal} {Phys. Rev. E}\ }\textbf {\bibinfo {volume} {84}},\ \bibinfo
  {pages} {056704} (\bibinfo {year} {2011})}\BibitemShut {NoStop}%
\bibitem [{\citenamefont {Lang}\ and\ \citenamefont
  {L\"auchli}(2019)}]{lang19}%
  \BibitemOpen
  \bibfield  {author} {\bibinfo {author} {\bibfnamefont {T.~C.}\ \bibnamefont
  {Lang}}\ and\ \bibinfo {author} {\bibfnamefont {A.~M.}\ \bibnamefont
  {L\"auchli}},\ }\bibfield  {title} {\bibinfo {title} {Quantum Monte Carlo
  Simulation of the Chiral Heisenberg Gross-Neveu-Yukawa Phase Transition with
  a Single Dirac Cone},\ }\href
  {https://doi.org/10.1103/PhysRevLett.123.137602} {\bibfield  {journal}
  {\bibinfo  {journal} {Phys. Rev. Lett.}\ }\textbf {\bibinfo {volume} {123}},\
  \bibinfo {pages} {137602} (\bibinfo {year} {2019})}\BibitemShut {NoStop}%
\bibitem [{\citenamefont {Hohenadler}\ \emph {et~al.}(2022)\citenamefont
  {Hohenadler}, \citenamefont {Liu}, \citenamefont {Sato}, \citenamefont
  {Wang}, \citenamefont {Guo},\ and\ \citenamefont {Assaad}}]{hohenadler22}%
  \BibitemOpen
  \bibfield  {author} {\bibinfo {author} {\bibfnamefont {M.}~\bibnamefont
  {Hohenadler}}, \bibinfo {author} {\bibfnamefont {Y.}~\bibnamefont {Liu}},
  \bibinfo {author} {\bibfnamefont {T.}~\bibnamefont {Sato}}, \bibinfo {author}
  {\bibfnamefont {Z.}~\bibnamefont {Wang}}, \bibinfo {author} {\bibfnamefont
  {W.}~\bibnamefont {Guo}},\ and\ \bibinfo {author} {\bibfnamefont {F.~F.}\
  \bibnamefont {Assaad}},\ }\bibfield  {title} {\bibinfo {title} {Thermodynamic
  and dynamical signatures of a quantum spin Hall insulator to superconductor
  transition},\ }\href {https://doi.org/10.1103/PhysRevB.106.024509} {\bibfield
   {journal} {\bibinfo  {journal} {Phys. Rev. B}\ }\textbf {\bibinfo {volume}
  {106}},\ \bibinfo {pages} {024509} (\bibinfo {year} {2022})}\BibitemShut
  {NoStop}%
\bibitem [{\citenamefont {Castro~Neto}\ \emph {et~al.}(2009)\citenamefont
  {Castro~Neto}, \citenamefont {Guinea}, \citenamefont {Peres}, \citenamefont
  {Novoselov},\ and\ \citenamefont {Geim}}]{castroneto09}%
  \BibitemOpen
  \bibfield  {author} {\bibinfo {author} {\bibfnamefont {A.~H.}\ \bibnamefont
  {Castro~Neto}}, \bibinfo {author} {\bibfnamefont {F.}~\bibnamefont {Guinea}},
  \bibinfo {author} {\bibfnamefont {N.~M.~R.}\ \bibnamefont {Peres}}, \bibinfo
  {author} {\bibfnamefont {K.~S.}\ \bibnamefont {Novoselov}},\ and\ \bibinfo
  {author} {\bibfnamefont {A.~K.}\ \bibnamefont {Geim}},\ }\bibfield  {title}
  {\bibinfo {title} {The electronic properties of graphene},\ }\href
  {https://doi.org/10.1103/RevModPhys.81.109} {\bibfield  {journal} {\bibinfo
  {journal} {Rev. Mod. Phys.}\ }\textbf {\bibinfo {volume} {81}},\ \bibinfo
  {pages} {109} (\bibinfo {year} {2009})}\BibitemShut {NoStop}%
\bibitem [{sup()}]{supplemental}%
  \BibitemOpen
  \bibinfo {note} {See Supplemental Material, which additionally includes
  Refs.~\cite{sun09,ray20}, for (i)~technical details of the quantum Monte
  Carlo simulations, (ii)~additional finite-size scaling analyses, (iii)~a
  discussion of the temperature dependence of the charge susceptibility in the
  QBT regime, and (iv)~additional finite-temperature Monte Carlo data for the
  charge susceptibility.}\BibitemShut {Stop}%
\bibitem [{\citenamefont {Assaad}\ \emph {et~al.}(2025)\citenamefont {Assaad},
  \citenamefont {Bercx}, \citenamefont {Goth}, \citenamefont {Götz},
  \citenamefont {Hofmann}, \citenamefont {Huffman}, \citenamefont {Liu},
  \citenamefont {Toldin}, \citenamefont {Portela},\ and\ \citenamefont
  {Schwab}}]{alf-doc}%
  \BibitemOpen
  \bibfield  {author} {\bibinfo {author} {\bibfnamefont {F.~F.}\ \bibnamefont
  {Assaad}}, \bibinfo {author} {\bibfnamefont {M.}~\bibnamefont {Bercx}},
  \bibinfo {author} {\bibfnamefont {F.}~\bibnamefont {Goth}}, \bibinfo {author}
  {\bibfnamefont {A.}~\bibnamefont {Götz}}, \bibinfo {author} {\bibfnamefont
  {J.~S.}\ \bibnamefont {Hofmann}}, \bibinfo {author} {\bibfnamefont
  {E.}~\bibnamefont {Huffman}}, \bibinfo {author} {\bibfnamefont
  {Z.}~\bibnamefont {Liu}}, \bibinfo {author} {\bibfnamefont {F.~P.}\
  \bibnamefont {Toldin}}, \bibinfo {author} {\bibfnamefont {J.~S.~E.}\
  \bibnamefont {Portela}},\ and\ \bibinfo {author} {\bibfnamefont
  {J.}~\bibnamefont {Schwab}},\ }\bibfield  {title} {\bibinfo {title} {{The ALF
  (Algorithms for Lattice Fermions) project release 2.4. Documentation for the
  auxiliary-field quantum Monte Carlo code}},\ }\href
  {https://doi.org/10.21468/SciPostPhysCodeb.1-v2.4} {\bibfield  {journal}
  {\bibinfo  {journal} {SciPost Phys. Codebases}\ }\textbf {\bibinfo {volume}
  {1-v2.4}} (\bibinfo {year} {2025})}\BibitemShut {NoStop}%
\bibitem [{\citenamefont {Beach}(2004)}]{beach04}%
  \BibitemOpen
  \bibfield  {author} {\bibinfo {author} {\bibfnamefont {K.~S.~D.}\
  \bibnamefont {Beach}},\ }\bibinfo {title} {Identifying the maximum entropy
  method as a special limit of stochastic analytic continuation},\ \Eprint
  {https://arxiv.org/abs/cond-mat/0403055} {arXiv:cond-mat/0403055}\BibitemShut
  {NoStop}%
\bibitem [{\citenamefont {Li}\ \emph {et~al.}(2009)\citenamefont {Li},
  \citenamefont {Luican},\ and\ \citenamefont {Andrei}}]{li09}%
  \BibitemOpen
  \bibfield  {author} {\bibinfo {author} {\bibfnamefont {G.}~\bibnamefont
  {Li}}, \bibinfo {author} {\bibfnamefont {A.}~\bibnamefont {Luican}},\ and\
  \bibinfo {author} {\bibfnamefont {E.~Y.}\ \bibnamefont {Andrei}},\ }\bibfield
   {title} {\bibinfo {title} {Scanning Tunneling Spectroscopy of Graphene on
  Graphite},\ }\href {https://doi.org/10.1103/PhysRevLett.102.176804}
  {\bibfield  {journal} {\bibinfo  {journal} {Phys. Rev. Lett.}\ }\textbf
  {\bibinfo {volume} {102}},\ \bibinfo {pages} {176804} (\bibinfo {year}
  {2009})}\BibitemShut {NoStop}%
\bibitem [{\citenamefont {Geim}\ and\ \citenamefont
  {Novoselov}(2007)}]{geim07}%
  \BibitemOpen
  \bibfield  {author} {\bibinfo {author} {\bibfnamefont {A.~K.}\ \bibnamefont
  {Geim}}\ and\ \bibinfo {author} {\bibfnamefont {K.~S.}\ \bibnamefont
  {Novoselov}},\ }\bibfield  {title} {\bibinfo {title} {The rise of graphene},\
  }\href {https://doi.org/10.1038/nmat1849} {\bibfield  {journal} {\bibinfo
  {journal} {Nat. Mater.}\ }\textbf {\bibinfo {volume} {6}},\ \bibinfo {pages}
  {183} (\bibinfo {year} {2007})}\BibitemShut {NoStop}%
\bibitem [{nhr()}]{nhr-alliance}%
  \BibitemOpen
  \bibinfo {note}
  {\href{http://www.nhr-verein.de/en/our-partners}{http://www.nhr-verein.de/en/our-partners}}\BibitemShut
  {NoStop}%
\bibitem [{\citenamefont {Sun}\ \emph {et~al.}(2009)\citenamefont {Sun},
  \citenamefont {Yao}, \citenamefont {Fradkin},\ and\ \citenamefont
  {Kivelson}}]{sun09}%
  \BibitemOpen
  \bibfield  {author} {\bibinfo {author} {\bibfnamefont {K.}~\bibnamefont
  {Sun}}, \bibinfo {author} {\bibfnamefont {H.}~\bibnamefont {Yao}}, \bibinfo
  {author} {\bibfnamefont {E.}~\bibnamefont {Fradkin}},\ and\ \bibinfo {author}
  {\bibfnamefont {S.~A.}\ \bibnamefont {Kivelson}},\ }\bibfield  {title}
  {\bibinfo {title} {Topological Insulators and Nematic Phases from Spontaneous
  Symmetry Breaking in 2D Fermi Systems with a Quadratic Band Crossing},\
  }\href {https://doi.org/10.1103/PhysRevLett.103.046811} {\bibfield  {journal}
  {\bibinfo  {journal} {Phys. Rev. Lett.}\ }\textbf {\bibinfo {volume} {103}},\
  \bibinfo {pages} {046811} (\bibinfo {year} {2009})}\BibitemShut {NoStop}%
\bibitem [{\citenamefont {Ray}\ \emph {et~al.}(2020)\citenamefont {Ray},
  \citenamefont {Vojta},\ and\ \citenamefont {Janssen}}]{ray20}%
  \BibitemOpen
  \bibfield  {author} {\bibinfo {author} {\bibfnamefont {S.}~\bibnamefont
  {Ray}}, \bibinfo {author} {\bibfnamefont {M.}~\bibnamefont {Vojta}},\ and\
  \bibinfo {author} {\bibfnamefont {L.}~\bibnamefont {Janssen}},\ }\bibfield
  {title} {\bibinfo {title} {Soluble fermionic quantum critical point in two
  dimensions},\ }\href {https://doi.org/10.1103/PhysRevB.102.081112} {\bibfield
   {journal} {\bibinfo  {journal} {Phys. Rev. B}\ }\textbf {\bibinfo {volume}
  {102}},\ \bibinfo {pages} {081112(R)} (\bibinfo {year} {2020})}\BibitemShut
  {NoStop}%
\end{thebibliography}%
